%% file: main.tex
\documentclass[pdflatex,sn-mathphys-num]{sn-jnl}

\usepackage{graphicx}%
\usepackage[skins]{tcolorbox}
\usepackage{multirow}%
\usepackage{amsmath,amssymb,amsfonts}%
\usepackage{amsthm}%
\usepackage{mathrsfs}%
\usepackage[title]{appendix}%
\usepackage{xcolor}%
\usepackage{textcomp}%
\usepackage{manyfoot}%
\usepackage{booktabs}%
\usepackage{algorithm}%
\usepackage{algorithmicx}%
\usepackage{algpseudocode}%
\usepackage{listings}%
\usepackage{tabularx}%
\usepackage{tikz}
\usetikzlibrary{shadows} 
\usepackage{caption}
\usepackage{makecell}   
\usepackage{float}

\usepackage{booktabs}
\usepackage{threeparttable}

\theoremstyle{thmstyleone}%
\theoremstyle{thmstyletwo}%

\theoremstyle{thmstylethree}%

\begin{document}



\title[Article Title]{ AI Standardized Patient Improves Human Conversations in Advanced Cancer Care
}


\author*[1]{\fnm{Kurtis} \sur{Haut}}\email{khaut@u.rochester.edu}
\equalcont{These authors contributed equally to this work.}

\author[1]{\fnm{Masam} \sur{Hasan}}\email{m.hasan@rochester.edu}
\equalcont{These authors contributed equally to this work.}

\author[2]{\fnm{Thomas} \sur{Carroll}}

\author[2]{\fnm{Ronald} \sur{Epstein}}

\author[1]{\fnm{Taylan} \sur{Sen}}

\author[1]{\fnm{Ehsan} \sur{Hoque}}\email{mehoque@cs.rochester.edu}

\affil*[1]{\orgdiv{Computer Science}, \orgname{University of Rochester}, \orgaddress{\street{2513 Wegmans Hall}, \city{Rochester}, \postcode{14627-0226}, \state{NY}, \country{United States}}}

\affil[2]{\orgdiv{Medical Center}, \orgname{University of Rochester}, \orgaddress{\street{601 Elmwood Avenue}, \city{Rochester}, \postcode{14642}, \state{NY}, \country{USA}}}

\input{sections/00_Abstract}




\maketitle

\input{sections/01_Introduction}
\label{sec1}
\input{sections/02_Results}
\label{sec2}
\input{sections/03_Disscussion}
\label{sec3}
\input{sections/04_Methods}\label{sec4}
\input{sections/05_data_release}
\input{sections/06_Author_Contributions}

\section*{Supplementary information}

\bmhead{Acknowledgments}

\bibliography{ref}

\input{sections/99_appendix}\label{sec5}
 
\end{document}

%% file: sections/00_Abstract.tex

\abstract{Serious illness communication (SIC) in end-of-life care faces challenges such as emotional stress, cultural barriers, and balancing hope with honesty. Despite its importance, one of the few available ways for clinicians to practice SIC is with standardized patients, which is expensive, time-consuming, and inflexible. In this paper, we present SOPHIE, an AI-powered standardized patient simulation and automated feedback system. SOPHIE combines large language models (LLMs),  a lifelike virtual avatar, and automated, personalized feedback based on clinical literature to provide remote, on-demand SIC training. In a randomized control study with healthcare students and professionals, SOPHIE users demonstrated significant improvement across three critical SIC domains: Empathize, Be Explicit, and Empower. These results suggest that AI-driven tools can enhance complex interpersonal communication skills, offering scalable, accessible solutions to address a critical gap in clinician education.}


%% file: sections/01_Introduction.tex
\section{Introduction}


Serious illness communication (SIC) in end-of-life care remains one of the most challenging aspects of clinical practice \cite{shilling2024let, pasricha2020use,paladino2023improving,myers2024simplifying}. These are high-stakes conversations where clinicians must navigate weighty issues, where a poorly chosen word could have lasting consequences on a patient’s final days and the memories their loved ones carry forward. Low-quality SIC has been associated with poor patient and family prognostic understanding \cite{hagerty2005communicating}, perceived lack of emotional support \cite{korsch1972doctor}, lower quality healthcare outcomes and higher costs \cite{ha2010doctor,riedl2017influence,stewart1995effective,begum2014doctor,centers2015national,oates2000impact,beck2002physician}. Communication with advanced-stage cancer patients specifically poses a variety of challenges, including: the volume and complexity of medical information, often fast-paced office visits, and the emotional burden of these life-changing conversations, for clinicians, patients, and their loved ones. Despite their extensive medical training, many physicians struggle to deliver difficult news effectively \cite{back2002communicating,gessesse2023exploring,bagley2023delivering}, often resulting in patient anxiety, misaligned treatment decisions, and reduced quality of care \cite{starcke2012decision,arnsten2009stress,gruneir2007people}. Also costly is the terms of expensive and potentially burdensome treatments as well as malpractice claims\cite{humphrey2022frequency}. The American Society of Clinical Oncology and the National Cancer Institute affirm that improving SIC is not only an ethical imperative but also a shared priority across healthcare, benefiting patients, their doctors and the broader medical system \cite{healingpatient}\cite{gilligan2017patient}.

The current state-of-the-art for communication training relies on human Standardized Patients (SPs), who provide realistic role portrayals and high-quality, actionable feedback. However, high quality SP-based training is limited by high costs, restricted demographic diversity, and the inherent fatigue and variability that result from repeatedly performing emotionally demanding roles \cite{bosse2015cost,gillette2017cost,menez2024strategies,everett2005recruitment}. These limitations make it difficult for clinicians—whose schedules are already constrained—to access on-demand, consistent training. 

AI-driven systems show promise in overcoming  the limitations of traditional standardized patient (SP)-based training by offering automated, scalable, reproducible, and personalized training tailored to a clinician’s schedule. Bowers et al. (2024) provide a comprehensive scoping review of AI-driven virtual patient frameworks designed to teach communication skills to healthcare students \cite{bowers2024artificial}. Complementarily, Wang et al. (2024) demonstrate that integrating AI with wearable sensors enables a more objective measurement of communication quality in real-patient settings \cite{wang2024commsense}. Additionally, Kearns et al. (2024) show that human-AI teaming with LLMs can foster more empathetic conversations, as clinicians articulate AI-generated suggestions to support compassionate care \cite{kearns2024bridging}. Despite promising developments, there is a need to establish whether AI systems can cultivate communication skills that clinicians can autonomously apply in practice. Rather than supporting clinicians with real-time AI prompts during patient encounters, there is a need for asynchronous AI-driven training that enables clinicians to deliver conversations that are effective, empathetic, and empowering—without reliance on AI in real-time clinical conversations.

To address this knowledge gap, we formed a collaboration between computer scientists and clinicians specialized in SIC training to create SOPHIE (Standardized Online Patient for Healthcare Interaction Education), a virtual standardized patient platform. SOPHIE is designed to simulate a patient with newly diagnosed, advanced lung cancer who is unaware that her condition is incurable. The avatar displays distress, seeks information, and shows resistance to accepting her prognosis. The case scenario requires clinicians to guide SOPHIE through difficult treatment decisions, ranging from comfort care to life-prolonging chemotherapy that may reduce quality of life. In collaboration with oncologists and palliative care physicians, we developed a highly realistic role requiring key SIC skills \cite{back2009mastering}. The team has extensive experience in developing standardized patient roles for medical education and communication research \cite{elias2017social, epstein2017effect, epstein2001improving, hoerger2013values}. As a training model, we adopted the MVP (Medical, Value, Planning) communication framework from the Advanced Communication Training (ACT) course from the University of Rochester Medical Center (URMC), which focuses on the 3 Es—Empathize, be Explicit, and Empower \cite{horowitz2020mvp}. SOPHIE’s clinical scenario challenges users to: 1) assess the patient’s readiness to receive serious news (empower), 2) clearly deliver that news (be explicit), and 3) provide emotional support (empathize). After each interactive, spoken encounter with SOPHIE, clinicians receive immediate, automated, and personalized feedback that emulates the guidance of an expert facilitator.

As a foundation to create an accurate cancer patient simulation, we first analyzed oncologist-patient dialogues from validated datasets \cite{sen2017modeling,epstein2017effect}, identifying recurring communication patterns. We then developed a schema-based dialogue manager \cite{kane2023managing} powered by large language models (LLMs) that balances conversational flexibility with clinical accuracy. SOPHIE integrates a lifelike virtual avatar, rendered with advanced 3D graphics, capable of engaging in real-time spoken dialogue while displaying appropriate emotional cues (Figure \ref{fig:emotions}). The system supports both local deployment and remote access via video conferencing platforms, ensuring accessibility across diverse clinical and educational settings. To provide structured feedback, we established quantitative metrics for three core SIC domains: Empowerment, Empathy, and Explicitness. This combination of realism and automated feedback enables healthcare professionals to repeatedly practice and refine critical SIC skills in a psychologically safe environment. We fine-tuned the technology through an initial qualitative clinical evaluation \cite{ali2021novel}, evaluated the feasibility of the idea via a controlled pilot study\cite{haut2023validating}. 

\begin{figure}
    \centering
    \includegraphics[width=\linewidth]{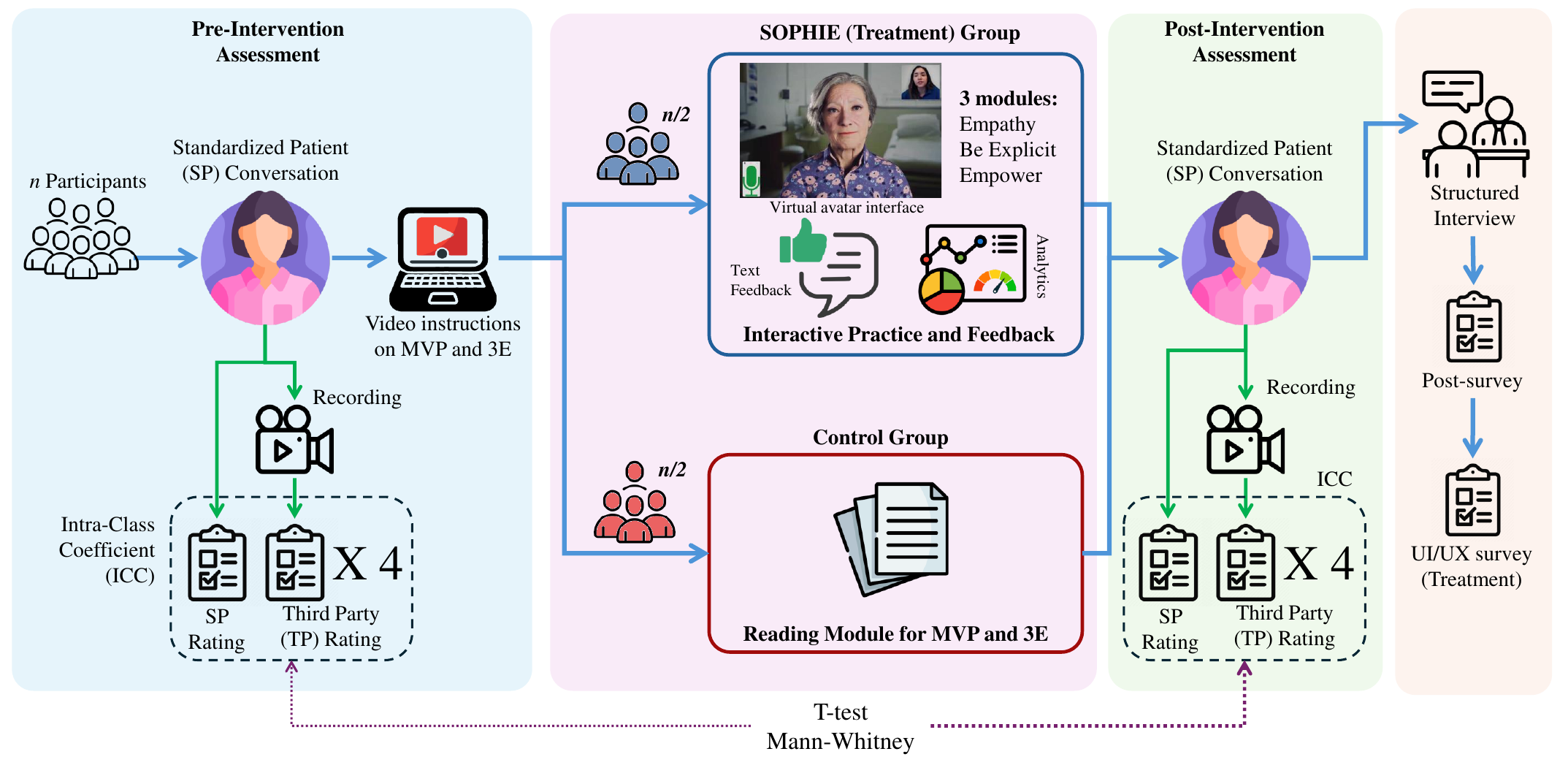}
    \caption{Randomized controlled study design. Participants were randomly assigned (blue arrows) to either the Control group, which completed a reading module on 3E communication skills, or the SOPHIE group, which received interactive practice with SOPHIE and personalized feedback. All participants engaged in standardized patient (SP) conversations both before and after the intervention, with evaluations provided by the SP and four third-party (TP) raters.}
    \label{fig:study-design}
\end{figure}

In this study, we conduct a randomized clinical trial of the AI-driven communication training intervention for health professionals. Figure \ref{fig:study-design} shows our randomized controlled study design. Participants were assigned to either an intervention group using the SOPHIE virtual patient system or a control group that received written study materials covering the same content. This study assessed SOPHIE’s impact on ``Three E's" Serious Illness Communication (SIC) skills based on the MVP model as rated by human standardized patients before and after the intervention, a gold standard for assessing communication skills. 
With a rigorous analysis, we measure the improvement in communication skills due to our intervention.

%% file: sections/02_Results.tex
\section{Results}

\subsection{Participant Characteristics}
Between June and December 2024, 51 participants were enrolled and assigned to either the SOPHIE intervention group (n = 26) or the control group (n = 25). No significant differences in demographic characteristics were observed between groups (see Table~\ref{tab:demographics}). A detailed breakdown of the characteristics of the study sample are summarized in Table~\ref{tab:dataset_summary}.

Participants were randomized using a stratified block approach based on clinical experience to ensure balanced allocation while allowing for natural demographic variation. To confirm the validity of the randomization process, we conducted a multivariate logistic regression analysis using participant demographic variables (age, gender, race, and professional background) as predictors of study arm assignment. None of the tested variables were found to be statistically associated with assignment to the SOPHIE or control group, indicating that group allocation was independent of demographic characteristics. This supports the internal validity of the study design and suggests that observed outcomes are unlikely to be driven by demographic confounds.

 All participants completed the demographic survey (N=51; Table \ref{tab:demographics}). A small portion of missing data was observed across follow-up measures due to technical issues (e.g., occasional Zoom recording failures) and participant attrition related to scheduling constraints. Specifically, 21 of 25 participants in the control arm and 22 of 26 in the SOPHIE arm completed the debrief survey. For the SOPHIE group, 23 of 26 completed the UI/UX survey. In total, 46 of 51 participants completed the post-study interview, while 506 of 510 conversation reviews were collected and rated. No formal dropouts were recorded.


\begin{table}[h]
    \centering
    \caption{Dataset Summary Statistics}
    \label{tab:dataset_summary}
    \begin{tabular}{ll}
        \toprule
        \textbf{Attribute} & \textbf{Count / Description} \\
        \midrule 
        Total Participants  & 51 \\
        \quad \textit{SOPHIE (Intervention) Arm} & 26 \\
        \quad \textit{Control Arm} & 25 \\
        Total Video Conversations & 102 \\
        Total Reviews & 506 \\
        Number of Standardized Patients (SP) & 13 \\
        Third-Party Reviewers (TP) & 4 \\
        Number of Reviews Per Video & 5 \\ 
        Unique Reviewers & 17 \\
        \midrule
        \textbf{Case Title Distribution} & \textbf{Count} \\
        \quad Case 1: Pat Smith (Stage IIIa lung cancer) & 168\\
        \quad Case 2: John/Lois Bell (Metastatic lung cancer) & 183  \\
        \quad Case 3: Jack/Jill Cooper (Post-nephrectomy for renal cell carcinoma) & 155 \\
        \bottomrule
    \end{tabular}
\end{table}

\begin{table}[h]
    \centering
    \caption{Participant Demographics}
    \label{tab:demographics}
    \begin{tabular}{lcc}
        \toprule
        \textbf{Category} & \textbf{Control (n=25)} & \textbf{SOPHIE (n=26)} \\
        \midrule
        \textbf{Age Distribution} & & \\
        \quad 18-24 & 9 & 14 \\
        \quad 25-34 & 12 & 6 \\
        \quad 35-44 & 1 & 1 \\
        \quad 45-54 & 2 & 2 \\
        \quad 55-64 & 0 & 1 \\
        \quad 65 or older & 1 & 2 \\
        \midrule
        \textbf{Gender} & & \\
        \quad Women & 72.0\% (18) & 84.6\% (22) \\
        \quad Men & 28.0\% (7) & 11.5\% (3) \\
        \quad Prefer not to say & 0\% & 3.8\% (1) \\
        \midrule
        \textbf{Race/Ethnicity} & & \\
        \quad White & 48.0\% (12) & 65.4\% (17) \\
        \quad Asian & 20.0\%  (5) & 23.1\% (6)\\
        \quad Black & 12.0\% (3) & 0.0\% (0) \\
        \quad Latino or Hispanic & 12.0 \% (3) & 3.8\% (1) \\
        \quad Middle Eastern / Other & 8.0\% (2) & 7.7\% (2) \\
        \midrule
        \textbf{Clinical Background} & & \\
        \quad Healthcare Professional Students & 14 & 14 \\
        \quad Practitioners (PAs, nurses, PhDs, and residents) & 11 & 12 \\
        \bottomrule
    \end{tabular}
\end{table}

\subsection{Primary Outcome}

To evaluate communication skill development, we analyzed changes in participants' scores (\textit{Empower}, \textit{Explicit}, \textit{Empathize}, and \textit{Overall}) between their first and second standardized patient (SP) conversations. Each conversation ($n = 102$) received five independent ratings—one from the participating SP and four from blinded third-party reviewers. Final scores were averaged across these ratings, and change scores ($\Delta$) were calculated by subtracting pre-intervention from post-intervention values. Inter-rater reliability was strong, with an intraclass correlation coefficient (ICC) of 0.882 (95\% CI: 0.82--0.93) among third-party reviewers (blinded to pre-/post- and intervention/control), confirming consistent evaluation.

\begin{figure}[h]
    \centering
    \includegraphics[width=\textwidth]{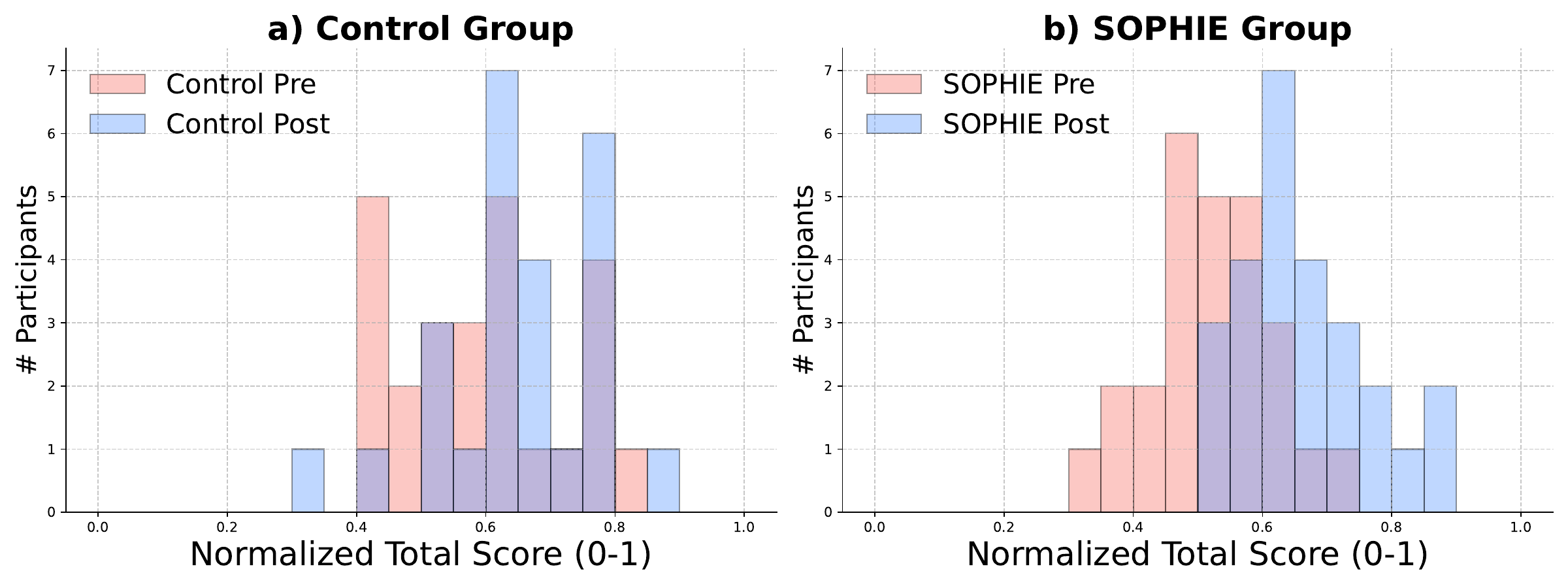}
    \caption{Histogram of Overall scores before (red) and after (blue) intervention for both control and SOPHIE groups (bin size $0.05$). The figure illustrates that the SOPHIE group showed a more pronounced rightward shift}
    \label{fig:score_distributions}
\end{figure}

\begin{figure}[htbp]
    \centering
    \includegraphics[width=0.8\textwidth]{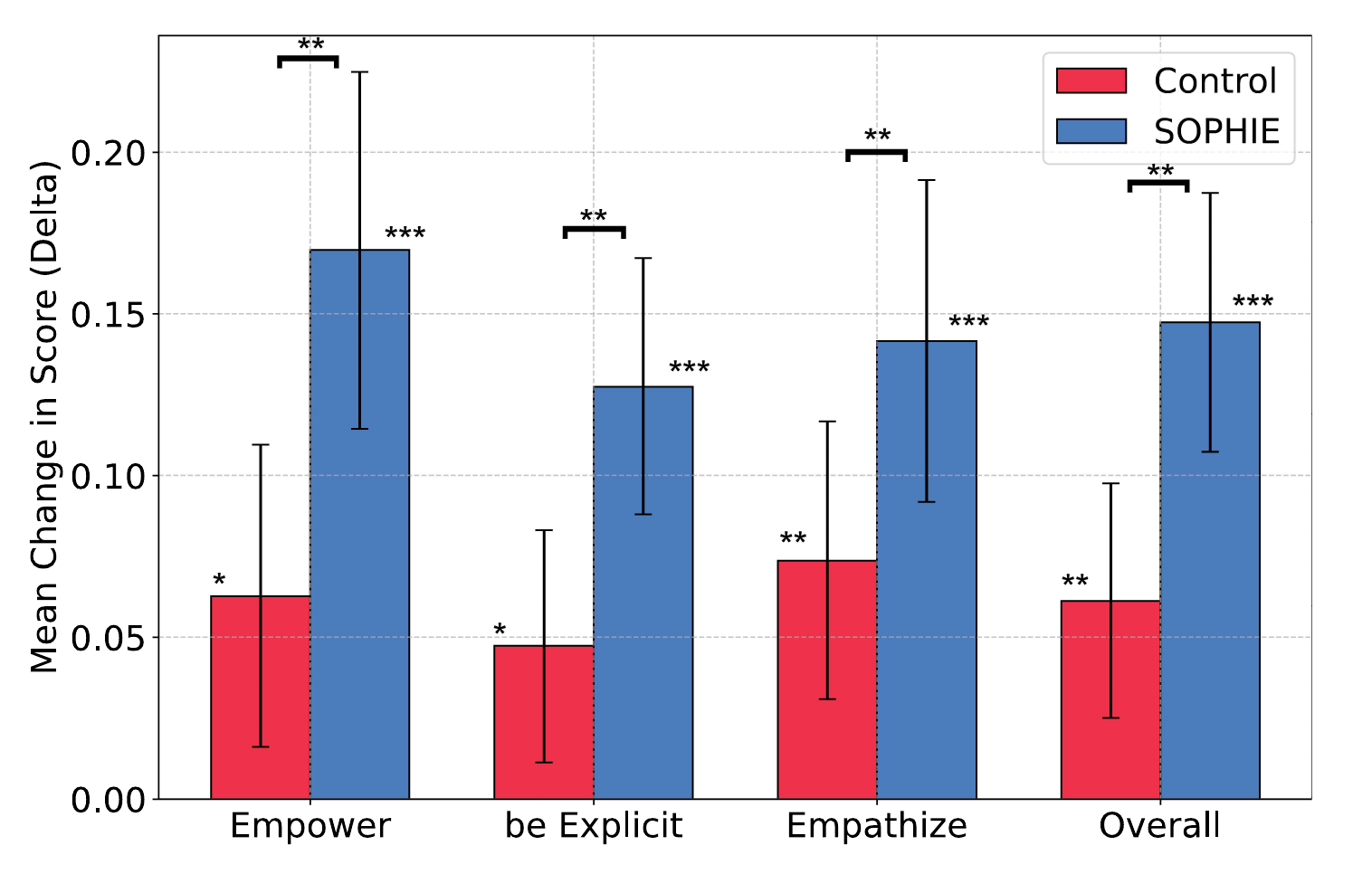}
    \caption{Comparison of communication skill improvement between the Control and SOPHIE groups. Bars represent the mean change ($\Delta$) in total score in 3E, with error bars indicating the 95\% confidence interval. Scores are normalized by minimum and maximum score, i.e., a 0.10 improvement represents a 10\% improvement in the maximum possible score. *, **, and *** indicates $p<0.05$, $p<0.01$, and $p<0.001$ respectively.}
    \label{fig:communication-skills}
\end{figure}

Both the control and SOPHIE intervention groups demonstrated significant improvement across all communication metrics after their second SP interaction (Table~\ref{tab:combined_arms}, Figure~\ref{fig:score_distributions}). However, the SOPHIE arm showed markedly greater gains. Participants receiving the SOPHIE intervention achieved larger $\Delta$ compared to the control in empowering patients  ($\Delta = 0.17$ vs $\Delta = 0.06$), being explicit ($\Delta = 0.13$ vs $\Delta = 0.05$)), and demonstrating empathy ($\Delta = 0.14$ vs $\Delta = 0.07$) (Figure~\ref{fig:communication-skills}). Table \ref{tab:combined_arms} shows that the SOPHIE group has a much larger effect size with much smaller $p values$.

Unpaired $t$-tests between Control and SOPHIE group $\Delta$ show significant improvements across Empower ($p = 0.004$, Cohens \textit{d} = $0.85$), Be Explicit ($p = 0.003$, Cohens \textit{d} = $0.87$), Empathy ($p = 0.002$, Cohens \textit{d} = $0.59$), and Overall ($p = 0.002$, Cohens \textit{d} = $0.92$). All 95\% confidence intervals for the pre-post differences are above zero, reinforcing the reliability of the findings (Table~\ref{tab:combined_arms}). Figure \ref{fig:communication-skills} demonstrates the improvement $\Delta$ of the Control and SOPHIE group and the 95\% Confidence Interval.

We observed a difference in the baseline skill of the SOPHIE and Control group (Table \ref{tab:combined_arms} Column Pre) and this difference was statistically significant in Empower ($p = 0.031$). To address baseline skill differences, we conducted a sensitivity analysis for Empower using a subset of participants with no measurable difference. The Results remained consistent with the full sample (Control: $\Delta = 0.08$; SOPHIE: $\Delta = 0.16$, $p=0.036$, Cohen's \textit{d} = $0.65$), supporting the robustness of the observed intervention effects (see Section \ref{sec:stat-analysis} in Methods for the analysis details).



\begin{table}[ht]
    \centering
    \caption{Communication Skill Improvement -- Control and SOPHIE Arms.}
    \label{tab:combined_arms}
    \begin{tabular}{lccccccc}
        \toprule
        \multicolumn{8}{c}{\textbf{Control Arm}} \\
        \midrule
        \multirow{2}{*}{\textbf{Skill}} & \multicolumn{2}{c}{\textbf{Pre}} & \multicolumn{2}{c}{\textbf{Post}} & \multirow{2}{*}{\textbf{$\Delta$ (95\% CI)}} & \multirow{2}{*}{\textbf{p-value}} & \multirow{2}{*}{\textbf{Cohen's d}} \\
        \cmidrule(lr){2-3} \cmidrule(lr){4-5}
        & \makecell{Mean} & \makecell{SD} & \makecell{Mean} & \makecell{SD} & & & \\
        \midrule
        Empower     & 0.49 & 0.14 & 0.55 & 0.13 & 0.06 (0.02--0.11) & 0.010 & 0.46 (Medium) \\
        Be Explicit & 0.71 & 0.11 & 0.75 & 0.11 & 0.05 (0.01--0.08) & 0.012 & 0.45 (Medium) \\
        Empathize   & 0.58 & 0.16 & 0.65 & 0.15 & 0.07 (0.03--0.12) & 0.002 & 0.47 (Medium) \\
        Overall         & 0.59 & 0.13 & 0.65 & 0.12 & 0.06 (0.03--0.10) & 0.002 & 0.48 (Medium) \\
        \midrule
        \multicolumn{8}{c}{\textbf{SOPHIE Arm}} \\
        \midrule
        \multirow{2}{*}{\textbf{Skill}} & \multicolumn{2}{c}{\textbf{Pre}} & \multicolumn{2}{c}{\textbf{Post}} & \multirow{2}{*}{\textbf{$\Delta$ (95\% CI)}} & \multirow{2}{*}{\textbf{p-value}} & \multirow{2}{*}{\textbf{Cohen's d}} \\
        \cmidrule(lr){2-3} \cmidrule(lr){4-5}
        & \makecell{Mean} & \makecell{SD} & \makecell{Mean} & \makecell{SD} & & & \\
        \midrule
        Empower     & 0.41 & 0.10 & 0.58 & 0.12 & 0.17 (0.11--0.22) & $0.000$ & 1.53 (Large) \\
        Be Explicit & 0.66 & 0.08 & 0.79 & 0.07 & 0.13 (0.09--0.17) & $0.000$ & 1.61 (Large) \\
        Empathize   & 0.51 & 0.14 & 0.65 & 0.12 & 0.14 (0.09--0.19) & $0.000$ & 1.07 (Large) \\
        Overall         & 0.52 & 0.09 & 0.67 & 0.10 & 0.15 (0.11--0.19) & $0.000$ & 1.55 (Large) \\
        \bottomrule
    \end{tabular}
\end{table}

\subsection{Secondary Outcomes}

\subsubsection{Learner Perceptions and System Efficacy}
Participants using the SOPHIE system reported markedly greater confidence in their communication skill development compared to controls on a 1-5 likert scale, with average perceived improvement scores of 4.31 (SOPHIE) vs. 3.83 (control) on a 5-point scale ($p < 0.01$). 

The training system itself received strong evaluations, particularly for its actionable feedback on communication skills—the highest-rated feature (mean: 4.7/5). Users also praised SOPHIE’s usability (mean: 4.2/5), though ratings for the virtual human’s realism were more moderate (3.4/5). (Table~\ref{tab:user_ratings}).

\subsubsection{Qualitative Feedback from Interviews}  
Qualitative feedback from post-session participant interviews revealed three dominant themes: the inherent difficulty of serious illness conversations, SOPHIE’s value as a training tool, and opportunities for system refinement.  

\textbf{The Challenge of High-Stakes Communication:}  
Participants universally acknowledged the emotional complexity of serious illness dialogues, with many expressing surprise at their own discomfort. One participant admitted, ``I felt nervous—even \textit{really, really bad}—practicing these conversations," while another emphasized, ``Doctors are human too—nobody is born an expert." 

\textbf{SOPHIE’s Training Value:}  
The system’s accessibility and actionable feedback were widely praised. Participants likened the experience to ``being trained in a facility by a specialist, but from my laptop," highlighting its convenience without compromising pedagogical rigor. The specificity of feedback—described as ``actionable guidance, not just generic encouragement"—enabled targeted skill development. Notably, the judgment-free environment empowered experimentation: ``I tried new strategies because there was no risk of hurting a real patient," one participant explained. Repeated practice translated to tangible confidence gains, with users reporting feeling ``more prepared for real conversations."

\textbf{Limitations and Future Directions:}  
While effective, SOPHIE’s realism faced critiques. Participants noted ``creepy" or ``uncanny" interactions due to limited emotional range, abrupt speech patterns, and lack of eye contact. Robotic reinforcement phrases (e.g., ``Thank you for allowing me to express my emotions") further reduced perceived authenticity. To address these gaps, users proposed enhancing multimodal analysis (facial expressions, tone, body language) and integrating game elements (performance scoring, competition, ect.).  

\textbf{Broader Implications:}  
Beyond healthcare, participants recognized the 3E framework’s potential in business, law, education, and personal relationships. 



\begin{table}[h]
    \centering
    \caption{Average Participant Scores on Post-study Survey (1-5)}
    \label{tab:user_ratings}
    \begin{tabular}{lcc}
        \toprule
        \textbf{System Attribute} & \textbf{SOPHIE} & \textbf{Control} \\
        \midrule
        Perceived skill improvement & 4.31 & 3.83 \\
        Communication feedback quality & 4.68 & - \\
        Usability & 4.22 & - \\
        Virtual human realism & 3.40 & - \\
        \bottomrule
    \end{tabular}
\end{table}

%% file: sections/03_Disscussion.tex
\section{Discussion}

\textbf{AI-Driven Training Enhances Skill Acquisition and Confidence - }
This randomized controlled trial demonstrates that AI-assisted training with SOPHIE yields significantly greater improvements in clinician communication skills compared to self-directed learning. While both groups showed progress, SOPHIE participants achieved stronger gains across all measured skills—empowering patients, clarity, and empathy (Figure~\ref{fig:communication-skills})—underscoring the efficacy of structured, feedback-driven AI training. Remarkably, these benefits emerged from sessions under 30 minutes, suggesting brief AI interventions can efficiently address time constraints while offering scalable skill development. 

Notably, confidence gains were 13\% higher in the SOPHIE arm, reinforcing that interactive practice enhances both competence, self-assurance and suggesting learners recognized their own progress. This echoes participant interviews, where many described feeling unprepared for the emotional demands of these high-stakes conversations—highlighting a critical gap in clinical training and validating the need for deliberate, risk-free practice. While long-term skill retention remains unmeasured, AI systems offer sustained reinforcement through repeatable, fatigue-free training—a crucial advantage for habit formation. Over time, such practice could reduce clinicians' cognitive load during stressful encounters, and help maintain these skill improvements.

\textbf{Feedback Quality Outshines Avatar Realism - }
Participants prioritized SOPHIE’s actionable feedback (rated 4.7/5) over the avatar’s visual realism (3.4/5), suggesting that the training’s value depends more on the quality of guidance than on photorealistic visuals (Table~\ref{tab:user_ratings}). While participants recognized emotions such as sadness and concern in  interactions (Figure~\ref{fig:emotions}), some noted that SOPHIE’s expressions felt ``uncanny.” This highlights a common design challenge: avatars that look too human can be unsettling, while more stylized expressions may actually make emotional cues clearer. Future systems may benefit from intentionally favoring exaggerated, easy-to-read expressions over striving for subtle, lifelike realism. The cross-domain applicability positions SOPHIE not just as a clinical tool but as a versatile platform for mastering emotionally charged communication beyond healthcare.

\textbf{Broadening Horizons: Scalability and Equity - }
SOPHIE’s framework extends beyond serious illness communication (SIC). Customizable avatars See (Figure~\ref{fig:different_races, genders, ages})in Appendix enable practice with diverse patient identities—including underrepresented demographics—potentially mitigating biases and improving culturally sensitive care. At scale, even modest skill gains across clinicians could reduce disparities in patient outcomes. Applications in conflict resolution, medical error discussions, and misinformation management further position AI training as a versatile clinical tool.

\textbf{AI as a Collaborative Catalyst - }
These findings provide much-needed empirical support for the growing hypothesis that AI can serve as an adaptive tutor, accelerating expertise development in complex interpersonal domains\cite{bereiter1993surpassing}. By delivering deliberate practice with immediate feedback\cite{ende1983feedback}\cite{molloy2012impact}\cite{boud2013feedback}\cite{thomas2011giving}, AI systems accelerate competency in interpersonal domains traditionally reliant on human mentors. A hybrid model—AI for foundational training, humans for nuanced refinement—may optimize learning while preserving the irreplaceable ``human touch.” As AI evolves, maintaining alignment with clinician needs and patient-centered values will ensure these tools amplify, rather than eclipse, the art of healing.

%% file: sections/04_Methods.tex
\section{Methods}
In the following methodology section, we first detail the design of SOPHIE, outlining the MVP framework and 3E skill classification system, dialogue manager, virtual avatar interface, and feedback generation process. We then describe our randomized controlled study, including participant recruitment, development of the scoring rubric, study procedures, standardized patient cases, module content, outcome measures, and statistical analysis methods used to evaluate SOPHIE’s effectiveness in communication training.

\subsection{System Design}
\subsubsection{MVP Framework and 3E Skill Classification}
\label{subsec:mvp}
The \textbf{MVP} framework \cite{mvp} is a structured yet flexible approach for serious illness conversations (SICs) that is supported by three essential communication skills -- termed as \textit{3E}'s:

\begin{enumerate}
    \item \textbf{Empower} – Ensure patient autonomy, tailor communication, and engage in shared decision-making.
    \item \textbf{Be Explicit} – Communicate clearly, directly, and concisely to avoid confusion and distress.
    \item \textbf{Empathize} – Validate and explore emotions, as desired/allowed by the patient, to support patients and families through difficult discussions.
\end{enumerate}

SOPHIE is built on the 3E framework—Empathize, be Explicit, and Empower—which underpins the system’s ability to precisely quantify and deliver actionable feedback on core communication behaviors.

Both for our dialogue generation and feedback, it is important for us to identify if a statement from the user presents one of the 3E skills. Figure \ref{fig:skill-classifier} in the appendeix, demonstrates our skill classifier system using a rule based text classifier  and a BERT \cite{bert} classifier trained with conversation data from \cite{haut2023validating}. If either models identifies a statement as Empathetic, Empowering, or Explicit, our system tags the user statement with the labeled skill.

\subsubsection{Dialogue Manager}
\label{subsec:dialogue-manager}

Recent advancements in large language models (LLMs) have enabled open-ended conversational simulations across diverse topics, including medical communication \cite{chatbot1,chatbot2,chatbot3}. However, LLMs face critical limitations in educational settings, such as generating factually inconsistent outputs (hallucination), losing conversational context (forgetting), and producing unpredictable responses, making it difficult to align dialogues with learning objectives.

In contrast, rule-based schema-driven dialogue systems provide precise control over conversational scope and pedagogical goals \cite{kane2023managing,lissa,aging,ali2021novel}. Yet, their rigid structure limits adaptability to unscripted scenarios, leading to repetitive responses and poor generalization.

To address these challenges, we developed a hybrid schema-guided LLM-based dialogue manager that combines the controllability of rule-based systems with the versatility of LLMs. The schema-based system identifies the conversation state using a skill classifier (Section \ref{subsec:mvp}) and suggests responses to the \texttt{gpt-3.5-turbo} LLM, guiding discussions toward learning objectives while maintaining natural flow. For out-of-domain conversations, the LLM responds directly. This hybrid approach ensures controlled, yet spontaneous, learner interactions.

\subsubsection{Virtual Avatar Interface}
\label{subsec:avatar}

\begin{figure}
    \centering
    \includegraphics[width=\linewidth]{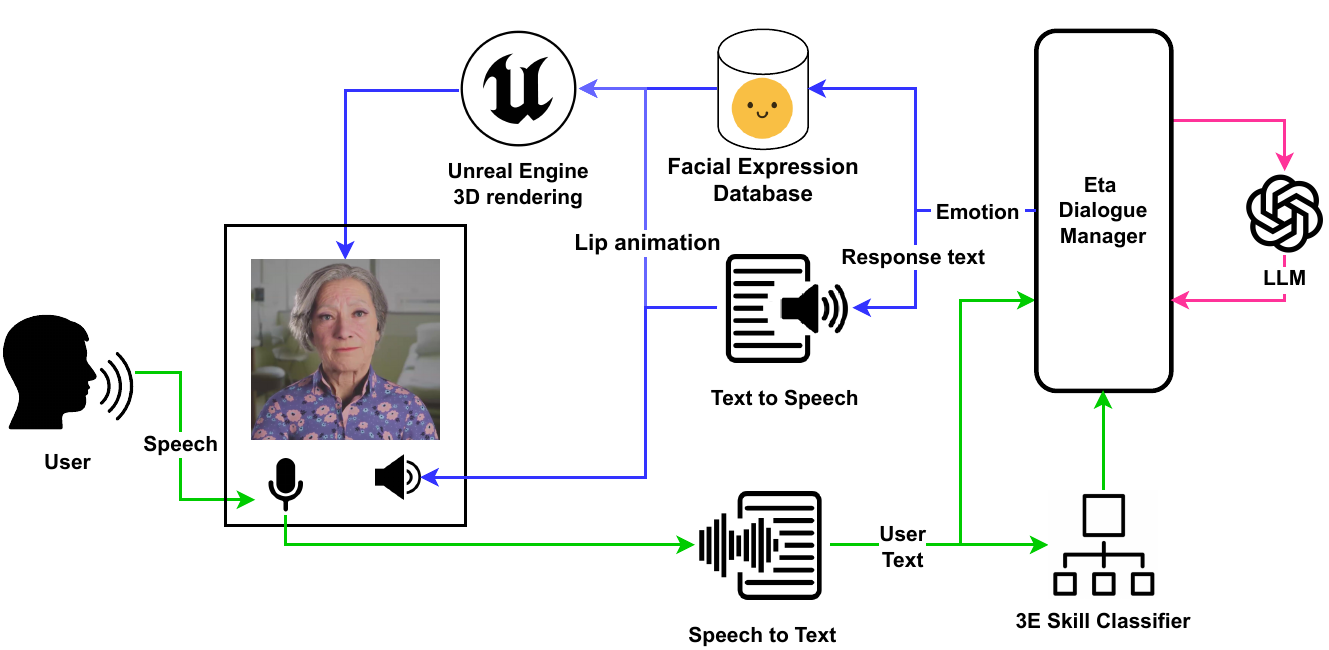}
    \captionsetup{justification=centerlast} 
    \caption{SOPHIE user speech to avatar response generation flow. Whole pipeline passes in nearly real-time.}
    \label{fig:sophie-flow}
\end{figure}

\begin{figure}[h!]
    \centering
    \begin{minipage}{0.3\textwidth}
        \centering
        \includegraphics[width=\linewidth]{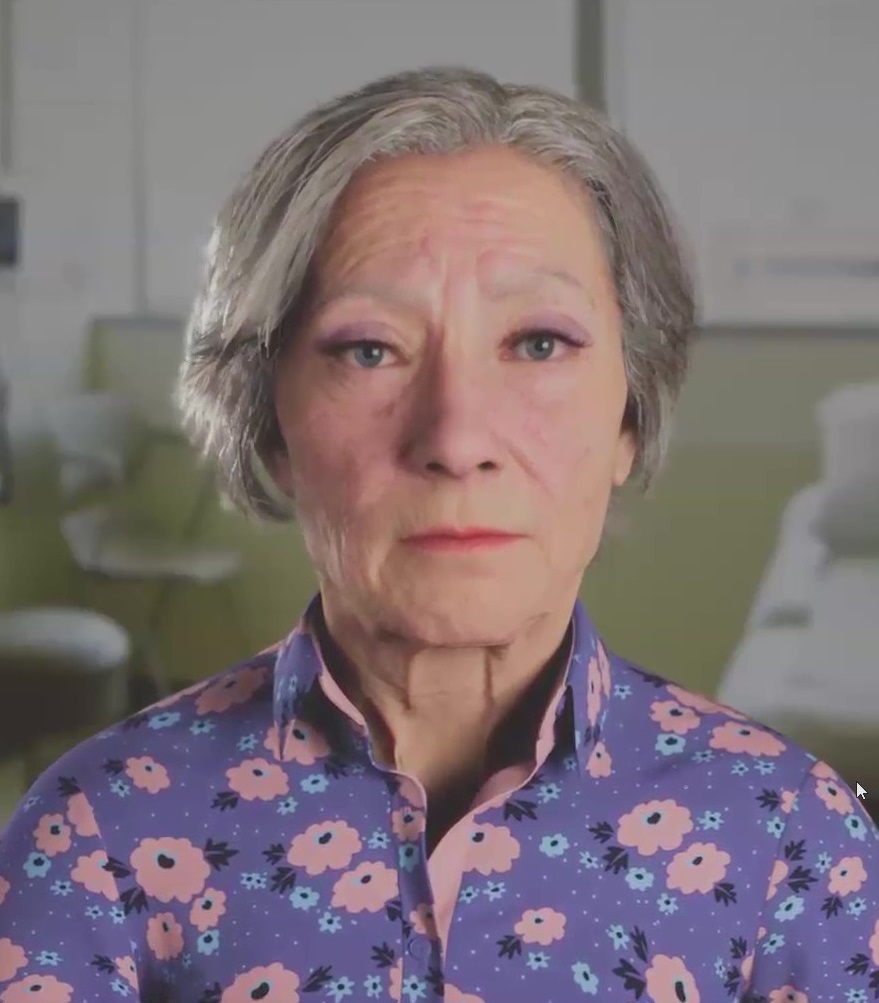}
        \caption*{(a) Neutral}
    \end{minipage}\hfill
    \begin{minipage}{0.3\textwidth}
        \centering
        \includegraphics[width=\linewidth]{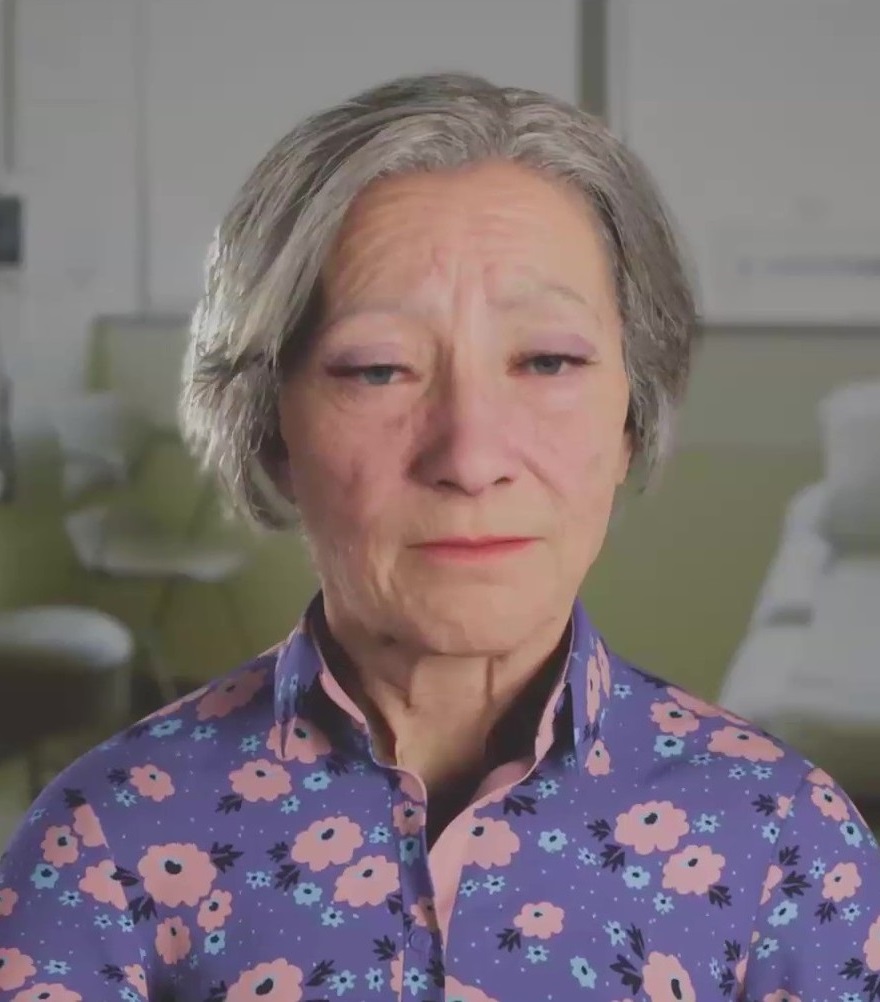}
        \caption*{(b) Sad}
    \end{minipage}\hfill
    \begin{minipage}{0.3\textwidth}
        \centering
        \includegraphics[width=\linewidth]{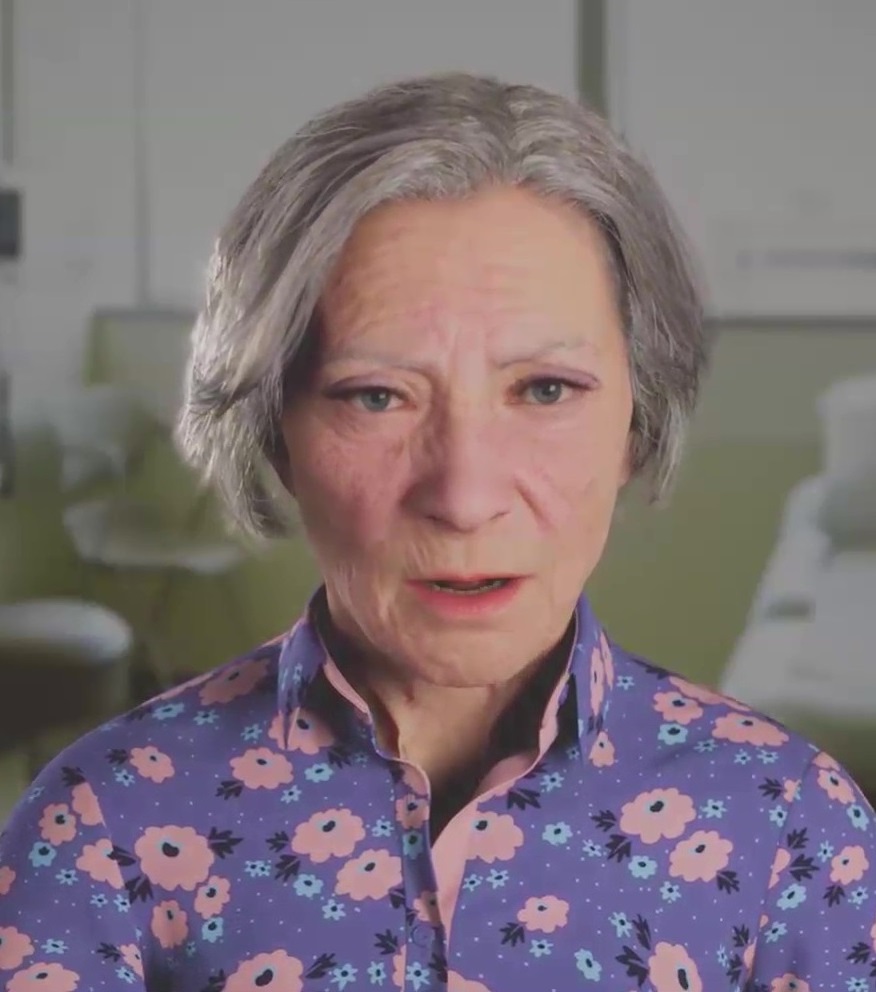}
        \caption*{(c) Angry}
    \end{minipage}
    \caption{Example of SOPHIE avatar facial expressions. Sad and angry are the most dominant emotions displayed in our conversation context. However, it is also capable of displaying Happy, Surprise, Fear, and Disgust.}
    \label{fig:emotions}
\end{figure}

Our system features SOPHIE, a high-fidelity, lifelike virtual patient—a 60-year-old woman with terminal lung cancer. SOPHIE engages users via local computer or video call. User speech is transcribed by a speech-to-text model and processed by the dialogue manager (Section \ref{subsec:dialogue-manager}), which generates a response and assigns an emotion tag. This tag governs SOPHIE’s facial expressions and vocal tone. A text-to-speech model then produces emotionally nuanced speech, synchronized with realistic lip movements and facial animations.

SOPHIE is built on Unreal Engine 5.1’s Metahuman platform \cite{metahuman}, enabling realistic body language and facial expressions. Real-time facial animations are driven by motion capture, with recordings for a neutral pose and six basic emotions (Happy, Sad, Angry, Surprised, Afraid, Disgusted). The dialogue manager selects animations based on the emotion tag. Figure \ref{fig:sophie-flow} illustrates the pipeline from user input to avatar response.

To ensure emotional clarity, facial and vocal cues prioritize expressiveness over naturalness. If a user fails to recognize Sophie’s emotional state, the intensity escalates. After three failures, the conversation terminates. This design fosters emotional engagement, enhancing the interaction’s realism and pedagogical effectiveness.

\subsubsection{Feedback}

\begin{figure}[H]  

\centering
\begin{minipage}{0.48\textwidth}
    \centering
    \includegraphics[width=\linewidth]{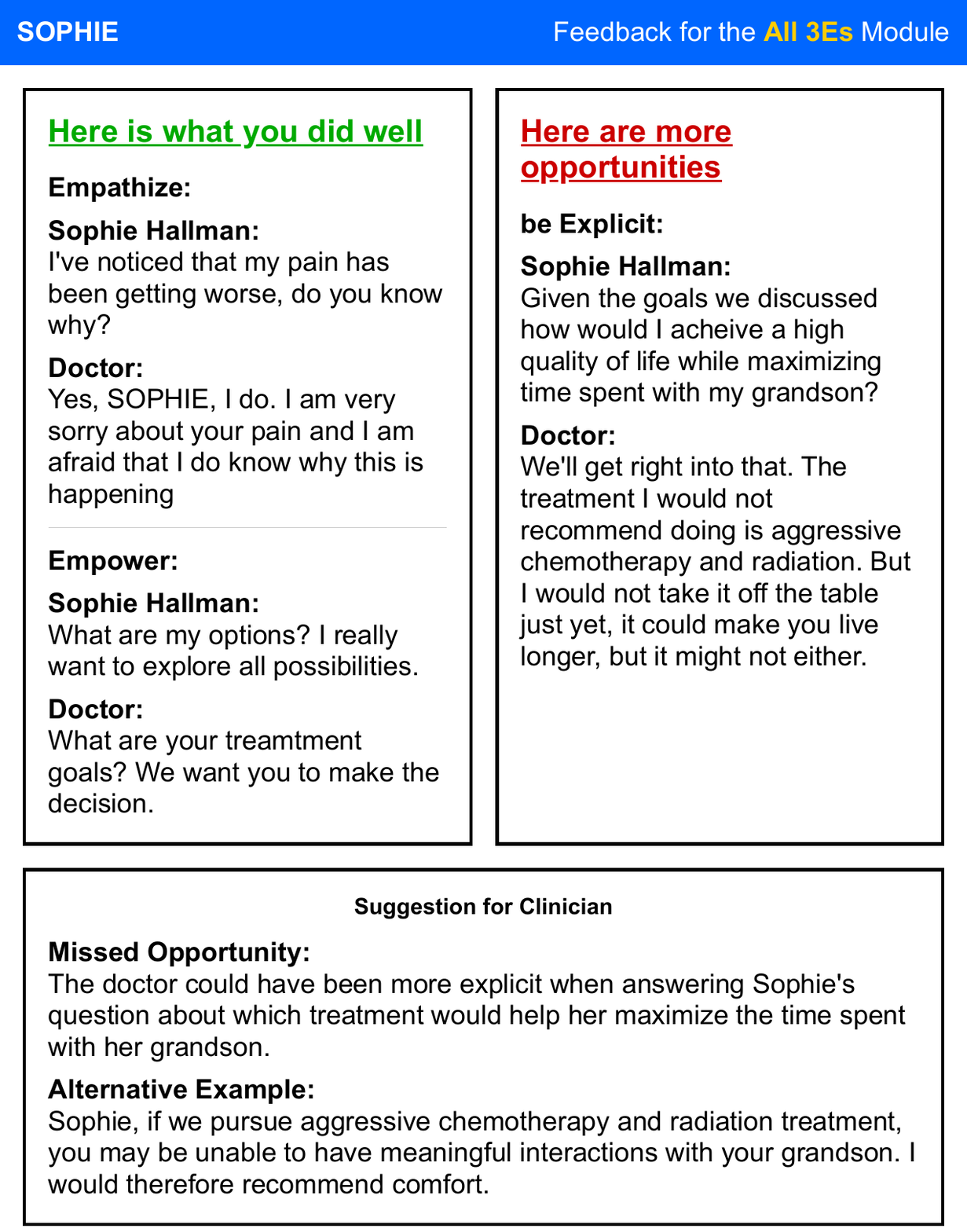}
    \caption*{(a) AI-generated descriptive and actionable feedback based on 3E}
\end{minipage}\hfill
\begin{minipage}{0.48\textwidth}
    \centering
    \includegraphics[width=\linewidth]{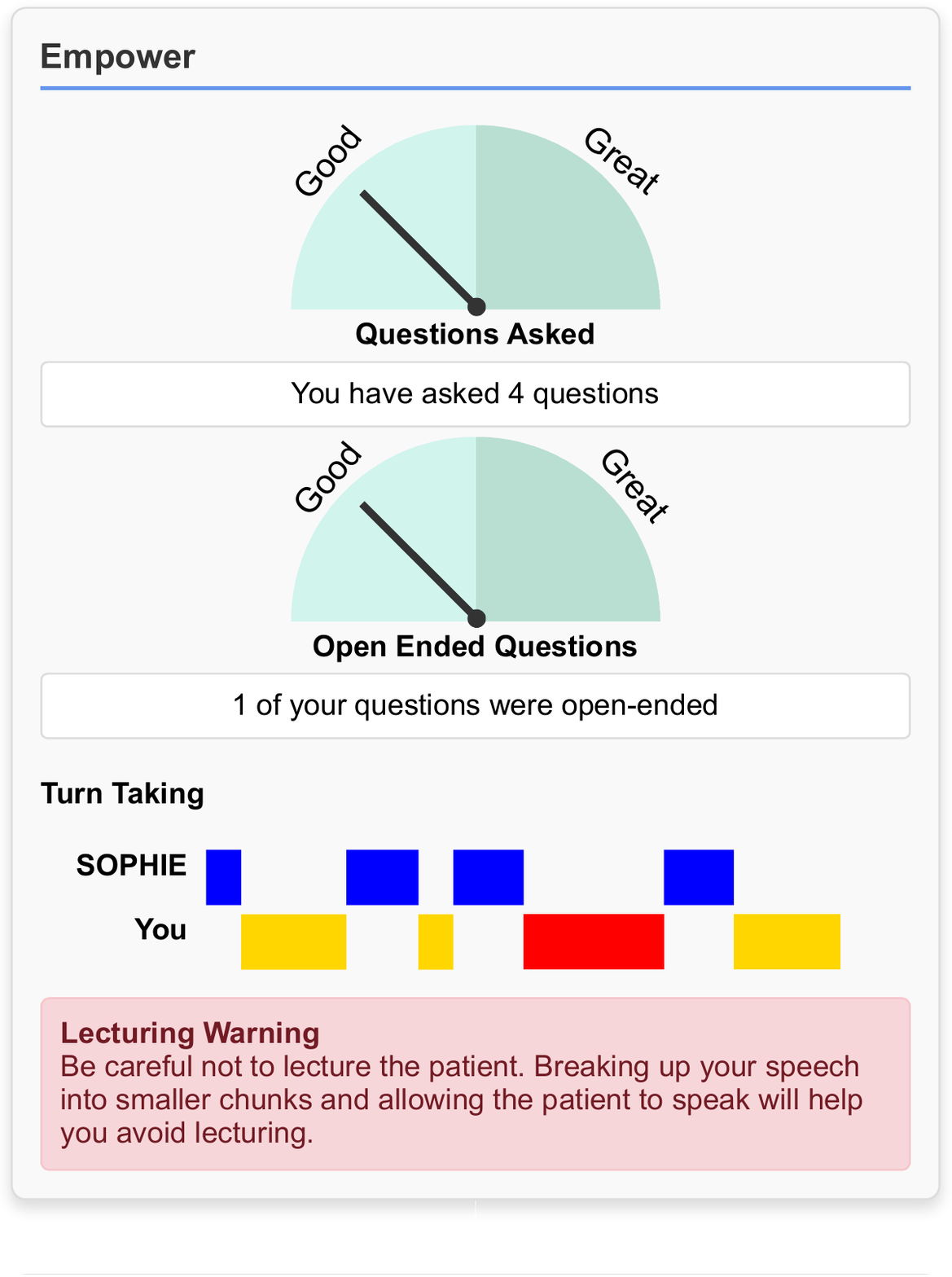}
    \caption*{(b) Empower Feedback}
\end{minipage}

\vspace{1em}

\begin{minipage}{0.48\textwidth}
    \centering
    \includegraphics[width=\linewidth]{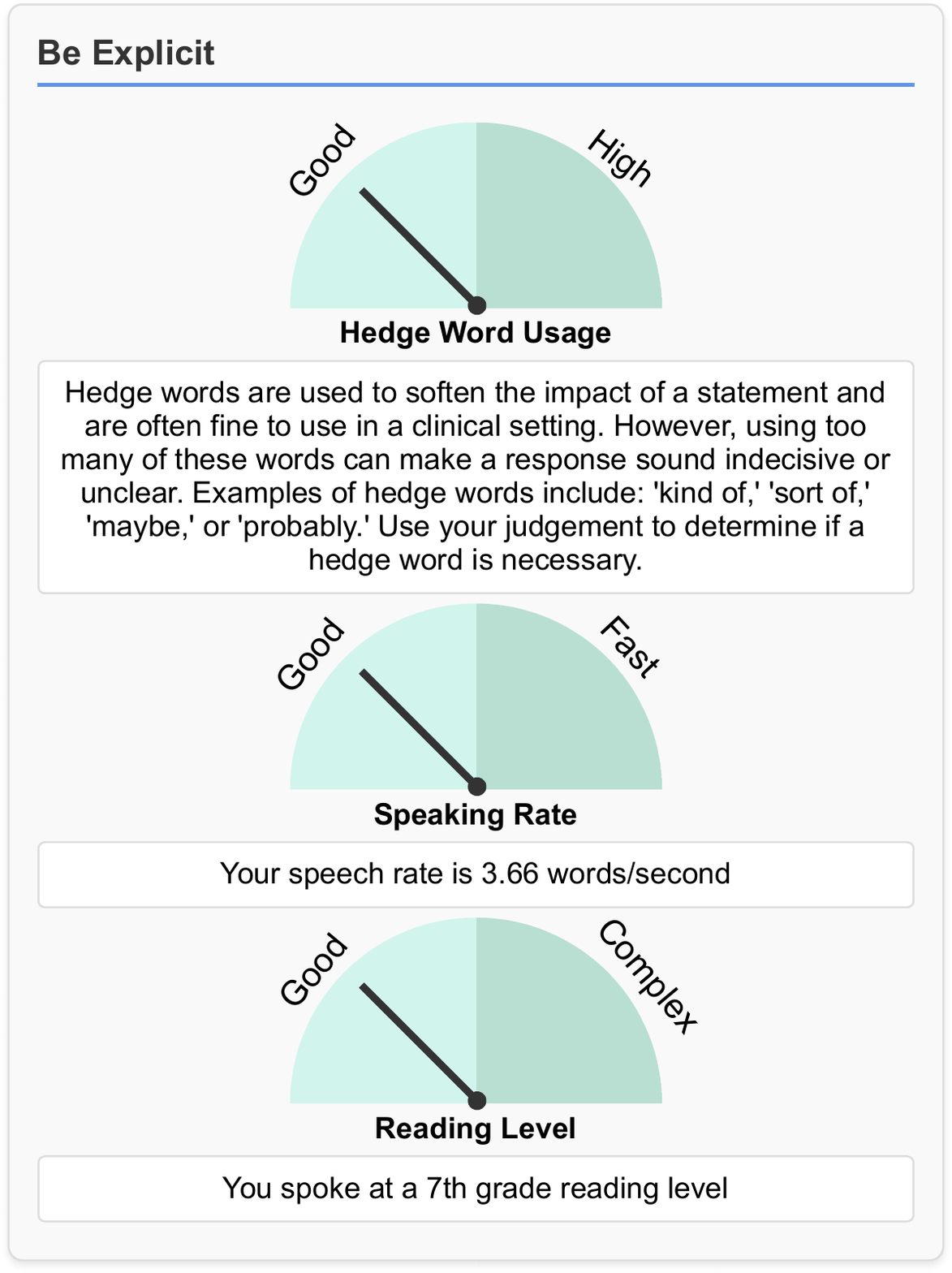}
    \caption*{(c) Be Explicit Feedback}
\end{minipage}\hfill
\begin{minipage}{0.48\textwidth}
    \centering
    \includegraphics[width=\linewidth]{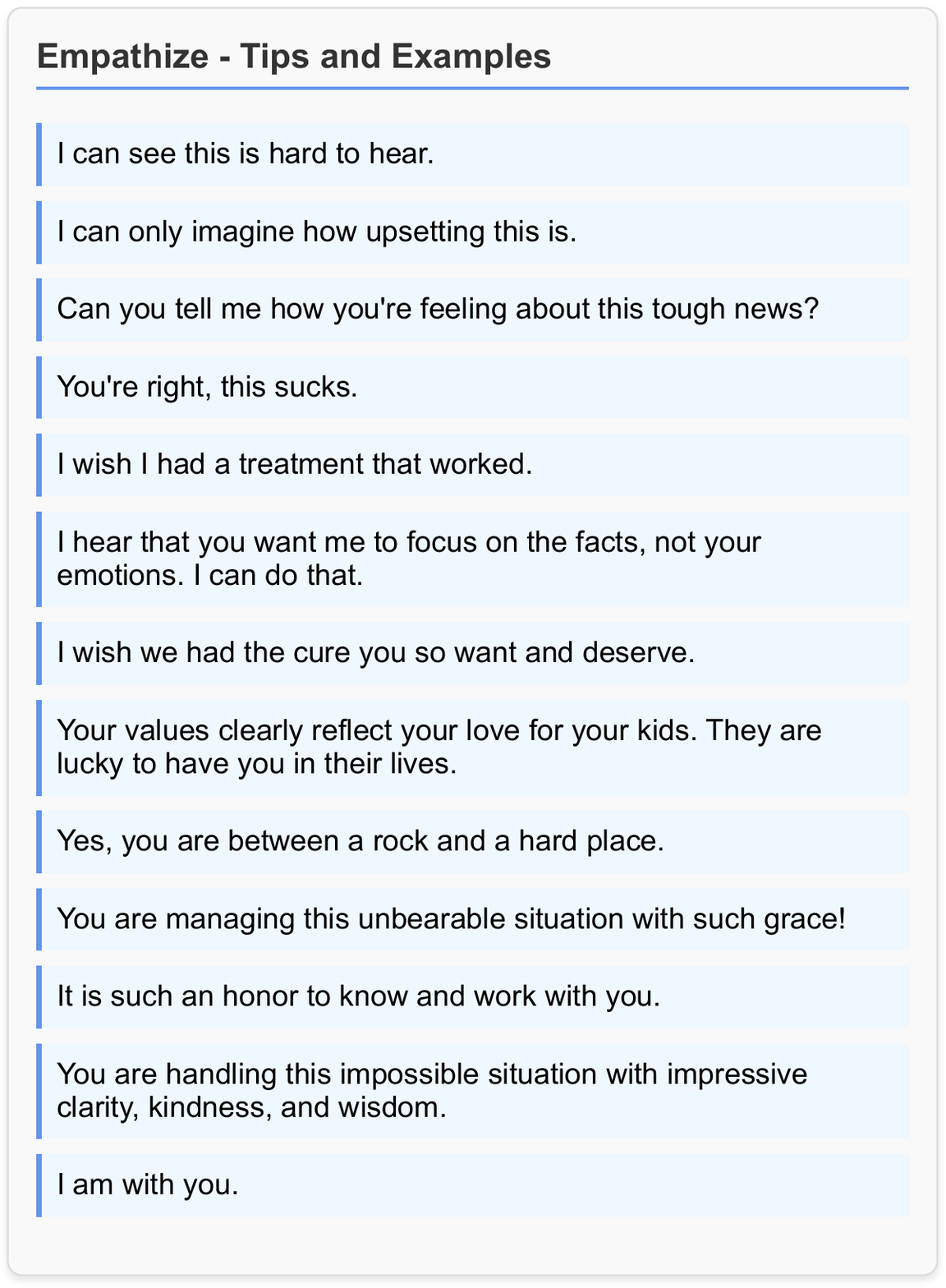}
    \caption*{(d) Empathy Feedback}
\end{minipage}

\caption{Displaying the SOPHIE feedback system. (a) shows the home page complete with AI-generated, actionable feedback. (b) to (d) show 3E-based feedback components which are available  at the request of the user as a downloadable pdf.}
\label{fig:feedback_new}
\end{figure}

SOPHIE’s feedback system has been iteratively refined based on input from palliative care experts, oncologists, and clinician user studies, including physicians, physician assistants, nurse practitioners, and medical and nursing students. Pilot study participants \cite{haut2023validating} emphasized the need for concise, actionable feedback, noting that earlier iterations with complex dials and gauges caused cognitive overload. Hence, we designed the interface to prioritize clarity and user control, featuring a ``What You Did Well" section (upper left) and an ``Opportunities to Improve" section (upper right). These sections focus on the application of the Empathize, Be Explicit, and Empower (3Es) skills, with interactive buttons allowing users to ``View Suggestion" for AI-driven recommendations or ``View Full Feedback" for a detailed breakdown (See Figure \ref{fig:feedback_new}). This ensures actionable insights for skill refinement.

At the core of SOPHIE’s feedback system is a custom dialogue manager that classifies user responses in real time. It identifies successful skill demonstrations for reinforcement in the ``What You Did Well" section and flags missed opportunities in the ``Opportunities to Improve" section (Figure \ref{fig:feedback_new} a) ). In ``Detailed Feedback", the user can see which of their spoken statement were classified into which skill (Appendix Figure \ref{fig:transcript}). The session length adapts dynamically to user performance: if a participant demonstrates a skill multiple times, the module ends to reinforce success; if the skill is not demonstrated, the system continues presenting opportunities within a structured loop. SOPHIE modulates emotional expressiveness to cue users, and if a participant repeatedly fails to apply the skill, the module ends after ~5 minutes, ensuring a focused training experience.

For personalized feedback, we employed \texttt{gpt-3.5-turbo-instruct} to analyze conversation transcripts and generate tailored suggestions. The LLM was prompted with the transcript, marked skill demonstrations, missed opportunities, instructional context from University courses (Namely, the Advanced Communication Training course), and few-shot examples for each 3E skill. This approach provides expert-like feedback, guiding users toward improvement. Additionally, selecting ``View Full Feedback" generates a downloadable PDF report with a quantified breakdown of the conversation, including the LLM-generated suggestion, a transcript with user statements mapped to SOPHIE’s responses, and skill demonstrations highlighted in green to reinforce positive behavior.

Pilot study insights guided the refinement of quantified feedback. Earlier empathy metrics (e.g., pronoun usage, sentiment trajectory) were discarded due to lack of contextual accuracy, replaced by concrete examples from the Advanced Communication Training (ACT) course at the University of Rochester Medical Center (URMC). For the Be Explicit module, well-received visualizations (e.g., hedge words, speaking rate, reading level) were retained. The Empower module continues to track question patterns, open-ended questions, and turn-taking dynamics, with excessive speaking flagged in red and accompanied by strategies for balanced dialogue. Both modules conclude with skill-building examples from the ACT course, reinforcing best practices for effective communication.


\subsection{Randomized Control Study}

\subsubsection{Participants}
Medical school, physician assistant (PA), and nursing students, as well as practitioners (PAs, nurses, PhDs, and residents) were recruited via email and in-person at the University of Rochester Medical Center (URMC). The study, approved by the institutional review board (IRB) as minimal-risk, was conducted remotely via Zoom. Each one-hour session included a three-way call with the participant, a standardized patient, and a researcher, with the intervention component limited to 30 minutes. All sessions were recorded, and participants provided written informed consent, including permission to share video recordings. Participants received a \$30 Amazon gift card as compensation, in addition to the educational benefits of the training. Even the control arm of the study was offered the opportunity to train with SOPHIE after all other study requirements were met.

\subsubsection{Scoring Rubrics to Measure Communication Skill} 
Assessing communication skills in medicine is inherently challenging due to their nuanced and context-dependent nature \cite{radziej2017assess,chauhan2023validated,stein2022general,murugesu2022challenges,rehim2017tools}. To address this, we developed a structured survey questionnaire and grading rubric evaluating the three core skills—Empathize, Be Explicit, and Empower (3Es)—through subsets of statements rated on a numeric scales.

For each of the 3E skills ($S$), we sum over the Likert scores associated with the skill and normalize them between 0 to 1 using Equation \ref{eq:normalized_score}. For the Overall score, questions related to all three skills are summed over. This process is repeated for rating provided by the Standardized Patient (SP) and the four Third-Party raters (TP), and averaged.


\begin{equation}
\text{Score}_S = \frac{1}{5} \sum_{i=\text{SP}, \text{TP}_{1-4}} \left( 
\frac{\sum_{j \in \{\text{question\_set}_S\}} \text{likert\_score\_value}_{j,i}}{\text{max\_points\_possible}_S} 
\right)
\label{eq:normalized_score}
\end{equation}

This multi-rater approach enhances objectivity by averaging subjective evaluations across five raters and the normalization approach makes the scores interpretable and comparable. The full set of statements used in the rubric is provided in Table \ref{tab:communication_questions} in the appendix.

\subsubsection{Study Design}
Participants were assessed on their baseline communication skills using our scoring rubric before undergoing their assigned intervention. Post-intervention, their skills were re-evaluated with the same rubric, enabling direct pre- and post-intervention comparisons.

We enrolled 51 participants, randomly assigned via stratified block randomization (based on level of clinical experience—practicing clinician vs. trainee) to either the SOPHIE intervention group (n=26) or control group (n=25). Both groups watched an instructional video on the 3Es communication framework. The SOPHIE group then completed three SOPHIE training modules on the 3Es, followed by a post-training conversation with the same standardized patient (SP) playing a role in a different case. The control group studied a paper on the MVP 3Es framework \cite{horowitz2020mvp} in lieu of SOPHIE training.

Three advanced cancer clinical scenarios, developed with URMC, were randomly assigned for each conversation. Conversations were rated by the SP and video-recorded via Zoom, with four third-party (TP) raters also evaluating the recordings. SPs were blinded to group assignment, and TP raters were blinded to both group assignment and conversation timing (pre- or post-intervention). Participants completed three forms of feedback collection: a demographic survey, a debrief survey, and a post-study structured interview. The demographic survey gathered information on participants’ age, gender, sex, race, ethnicity, childhood social class, marital status, parental status, languages spoken, and prior clinical experience (including type and years of experience). The debrief survey focused on participants’ perceived improvement in their communication skills after using SOPHIE, their willingness to engage in future modules, and their self-reported stress level, all assessed using a 5-point Likert scale. Following the debrief, participants took part in a 10-minute structured interview immediately after their second standardized patient (SP) interaction. Interview questions explored learning outcomes, emotional responses, key takeaways, and retention of the 3 Es framework (Empathize, be Explicit, Empower). Additionally, participants provided targeted UI/UX feedback, including evaluations of system usability, emotional realism of the virtual human, the conversational realism of SOPHIE’s dialogue responses, and the perceived value of various user interface elements. Please see Tables \ref{tab:shared_questions}, \ref{tab:arm_specific_questions}, \ref{tab:interview_shared}, \ref{tab:UIUX_survey1} and \ref{tab:UIUX_survey2} in Appendix.

\subsubsection{Standardized Patient Cases}
Participants engaged in simulated clinical encounters reflecting high-stakes, end-of-life discussions in advanced cancer scenarios. Developed in collaboration with palliative care specialists and oncologists at the University of Rochester Medical Center (URMC) as part of the Advanced Communication Training (ACT) workshop, these cases required participants to deliver serious news, discuss prognosis, and explore treatment options.

Three cases were randomly assigned for each interaction:
\begin{itemize}
\item \textbf{Pat Smith}: A patient with Stage IIIa lung cancer, hopeful for a cure after aggressive chemotherapy and radiation, now facing terminal metastatic disease.
\item \textbf{John/Lois Bell}: A patient with metastatic lung cancer who avoided treatment until now, now learning the cancer is rapidly advancing and incurable.
\item \textbf{Jack/Jill Cooper}: A patient post-nephrectomy for renal cell carcinoma, now presenting with extensive metastases and no available targeted therapies, signaling a terminal prognosis.
\end{itemize}

We partnered with Professional Communication Simulators \footnote{\url{https://www.professionalcommunicationsimulators.com}} to select 13 professional Standardized Patients (SPs) based on availability. SPs were trained in the 3 Es framework and rating scale (see Table~\ref{tab:communication_questions} in the Appendix) during a workshop led by the ACT program director at URMC. The session was recorded and subsequently made available for SPs to revisit as needed for reinforcement. Additionally, four third-party (TP) reviewers, selected from the ACT course at URMC, independently evaluated video recordings of the encounters.

\subsubsection{Module Content}
The SOPHIE modules were developed based on scenarios adapted from prior validated studies, including the SIP study \cite{srinivasan2006connoisseurs} and the VOICE study \cite{epstein2017effect}, both of which featured standardized patient roles rated as highly realistic by experienced clinicians. SOPHIE’s clinical scenario and dialogue drew directly from these evidence-based frameworks, ensuring alignment with real-world communication challenges encountered in serious illness communication (SIC). The virtual patient was presented as an older adult receiving scan results confirming stage IV metastatic cancer, with a prognosis of six months to one year without treatment and one to two years with treatment likely to involve significant side-effects.

SOPHIE’s emotional tone was designed to vary contextually, shifting between neutral, sadness, worry, and frustration depending on the conversational flow. The system was programmed to escalate SOPHIE’s emotional distress when participants failed to address her affective state using strategies from the 3 Es framework. This emotionally responsive behavior was intended to reflect the complexity of real patient interactions and support the transfer of learned skills into clinical practice.

Participants completed all modules in a single session, with each module consisting of 3–5 minutes of spoken conversation with SOPHIE, followed by 3–5 minutes of feedback review. The entire session was capped at 30 minutes. See Figure \ref{fig:study-design} for how the study design flow works.


\begin{itemize}
\item \textbf{Empathize Module}: Participants practiced acknowledging SOPHIE’s emotional distress, validating her feelings, and offering emotional support as she processed the news.
\item \textbf{Be Explicit Module}: Participants focused on delivering the prognosis, including uncertainty, clearly and concisely, avoiding vagueness or overly long responses that could lead to misunderstandings.
\item \textbf{Empower Module}: Participants engaged SOPHIE in discussions about her values, explored her treatment preferences, and guided her in developing a patient-centered care plan.
\end{itemize}

Across all modules, participants received personalized feedback highlighting effective use of the targeted skill and missed opportunities and could receive detailed, quantified feedback if desired (See Figure \ref{fig:feedback_new}).

Control participants engaged with a reading module on the MVP (Medical Situation, Values, and Plan) and 3Es (Empathize, Be Explicit, Empower) pedagogical framework \cite{horowitz2020mvp}. Like the SOPHIE group, they were given a maximum of 30 minutes to complete the reading. However, the control module did not include interactive dialogue practice or personalized feedback. Control participants verbally confirmed reading the paper to the research proctor but were not quizzed on comprehension or asked to reflect on their learning.

\subsubsection{Study Outcomes}
The primary outcome was the mean change in communication skill scores, evaluated using the 3E framework (Empathize, Be Explicit, Empower) scoring rubric. Pre- and post-intervention scores were compared within each study arm using paired t-tests, and across arms using an unpaired t-test to determine if the SOPHIE group showed significantly greater improvement than the control group.

Secondary outcomes included user experience metrics, assessed through post-session UI/UX surveys and interviews. These captured participants’ perceptions of usability, realism of the virtual patient interaction, and quality of system feedback. Survey responses were analyzed quantitatively, with Likert-scale ratings summarized as means and standard deviations. Thematic analysis of interview transcripts was conducted by the research team to identify key themes, limitations, and potential system improvements.

\subsubsection{Statistical Analysis}
\label{sec:stat-analysis}
The target sample size was determined via a power analysis using pilot data, which showed an effect size of 0.82 (Cohen’s d). A one-sided, two-sample  t-test at a 5\% Type I error rate was selected to detect improvement in the SOPHIE arm relative to the control. To achieve at least 80\% power, a minimum of 48 participants (24 per arm) was planned, with an option to increase to 60 (30 per arm) if non-normality necessitated a Mann-Whitney test.

Participants were rated pre-intervention and post-intervention using the 3E scoring rubric, generating two summary scores per participant. Within-group changes were evaluated using paired t-tests, while between-group differences were assessed using an unpaired t-test on the pre-post score deltas. Effect sizes were calculated using Cohen’s d. Data distribution approximated normality, allowing the study to conclude at ~25 participants per arm. Analyses were conducted using Python statistical libraries, with significance set at $P<0.05$.

Inter-rater agreement was evaluated using the intraclass correlation coefficient (ICC). Four third-party (TP) reviewers rated 102 videos, with QSUM scores (sum of 18 questions rated 1–5) showing good consistency (ICC = 0.882, 95\% CI [0.82, 0.93]). Standardized Patients (SPs) were excluded from ICC calculations as they did not rate the same participants, but Mann-Whitney tests confirmed no significant differences in SP scores.

To assess whether participant demographic characteristics were associated with study arm assignment, we conducted a multivariate logistic regression with study arm (SOPHIE vs. Control) as the dependent variable. Independent variables included gender, age group (categorized as 18–34 vs. 35+), race, and participant background (clinician vs. student). Categorical variables were dummy encoded, and missing values were excluded listwise. The analysis revealed no significant associations between demographic variables and arm assignment, indicating that randomization successfully balanced the groups on measured baseline characteristics

\textbf{Sensitivity analysis:}To assess whether baseline differences in communication skill could explain the observed intervention effects, we conducted independent two-sample t-tests on pre-intervention scores for each of the three targeted domains: Empathize, be Explicit, and Empower. While no statistically significant differences were observed for Empathy ($p = 0.11$) and Explicitness ($p = 0.10$), Empower showed a statistically significant difference at baseline ($p = 0.03$), with the Control group starting higher on average. This raised the possibility that group differences in skill improvement were influenced by differing opportunities for growth. To address this, we performed a sensitivity analysis by restricting the sample to participants with overlapping pre-intervention score ranges. Specifically, we excluded participants in the Control group whose scores exceeded the highest baseline SOPHIE scores ($Empower > 0.645$), and those in the SOPHIE group whose scores fell below the lowest baseline Control scores ($Empower < 0.30)$. The resulting subset (Control: $n = 20$; SOPHIE: $n = 24$) showed no detectable differences in pre-intervention distributions ($p = 0.79$). We tracked these participants and measured their improvement $\Delta$.

%% file: sections/05_data_release.tex
\section*{Code and Data Release}
Alongside the paper, we are releasing the 506 ratings provided by professional raters, along with our analysis code at \url{https://github.com/ROC-HCI/SOPHIE-1.0}. The videos and timestamped transcripts will be available to researchers on a case-by-case basis, contingent on IRB approval.

%% file: sections/06_Author_Contributions.tex
\section*{Author Contributions}
\textbf{K.G.H.} led the SOPHIE project, including experiment conceptualization, methodology, project administration, investigation, and data curation. He also contributed to software iteration planning, formal analysis (including thematic analysis of interviews), visualization, and led writing—original draft and writing—review and editing. \textbf{M.H.} contributed to software development (including the virtual avatar, emotion system, and LLM integration), formal analysis (UI/UX survey and statistical validation), visualization, and writing—review and editing. He also led the restructuring of the manuscript and refined the technical description of the system architecture. \textbf{T.C.} was a co-principal investigator contributed to idea conceptualization, methodology, investigation, and resources. As the primary expert on the 3E communication framework, he ensured integration of realistic clinical content and contributed to writing—review and editing. \textbf{R.E.} was a co-principal investigator and contributed to idea conceptualization, methodology, and writing—review and editing. He co-led the design of the virtual standardized patient’s persona and the feedback delivery system with T.C. \textbf{T.S.} performed formal analysis, including statistical modeling and hypothesis testing, and contributed to data curation, visualization, and writing—review and editing. \textbf{E.H.} was the principal investigator of the project and contributed to the original idea conceptualization and study design for SOPHIE. He supervised the entire project, including the lead authors, K.G.H. and M.H., and provided strategic oversight throughout its development and deployment, and writing—review and editing.

%% file: sections/99_appendix.tex
\newpage
\section*{Appendix}

\begin{table}[h]
    \centering
    \begin{tabular}{p{1cm} p{13cm}} 
        \toprule
        \textbf{\#} & \textbf{Communication Skill Evaluation Questions} \\
        \midrule
        \multicolumn{2}{l}{\hspace{.5cm} \textbf{Empower Questions}} \\ 
        1  & The participant elicited the patient’s major concerns within the first 5 minutes of the conversation. \\
        2  & The participant asked for permission to share information about prognosis. \\
        3  & The participant asked how much information the patient would like concerning prognosis. \\
        4  & The participant checked the patient’s prognostic understanding by asking them to state what they understood, using a teach-back approach. \\
        5  & The participant actively encouraged the patient to ask questions using facilitative questions/statements (e.g., "What questions do you have?" or "At this point, many patients have questions..."). \\
        6  & The participant gave the SP multiple (2 or 3 or more) opportunities to talk. \\
        7  & Overall, I thought the participant was empowering. \\
        \midrule
        \multicolumn{2}{l}{\hspace{.5cm} \textbf{Be Explicit Questions}} \\ 
        8  & The participant described the medical situation (the cancer has spread) clearly and without euphemism or jargon. \\
        9  & The participant shared the prognosis accurately (a few weeks, a few months - less than one year). \\
        10 & The participant used clear language without euphemism or jargon when sharing the prognosis. \\
        11 & The participant used difficult-to-understand medical jargon. (Inverted during analysis) \\
        12 & The participant did a good job NOT lecturing the patient (uninterrupted information for what seemed like a long time). \\
        13 & Overall, I thought the participant was being explicit. \\
        \midrule
        \multicolumn{2}{l}{\hspace{.5cm} \textbf{Empathize Questions}} \\ 
        14 & The participant was generally empathetic. \\
        15 & The participant's language was empathetic. \\
        16 & The participant used silence appropriately in response to patient emotion. \\
        17 & The participant validated the SP's emotional responses. \\
        18 & Overall, I thought the participant aligned with the patient's emotional state. \\
        \bottomrule
    \end{tabular}
    \caption{Evaluation questions used to assess participant communication skills across the Empower, Be Explicit, and Empathize question sets. Overall questions (7,13, 18) are on a 1-10 scale, and the remaining questions are on a 1-5 scale.}
    \label{tab:communication_questions}
\end{table}

\begin{table}[htbp]
\centering
\caption{Shared Participant Survey Questions (Control and SOPHIE Arms)}
\label{tab:shared_questions}
\begin{tabular}{p{14cm}}
\hline
\textbf{Perceived Improvement of Communication Skills} \\
\hline
Please rate the following statements (1-5) with 5 representing strong agreement:
\begin{enumerate}
    \item I felt genuine empathy for the standardized patient acting out the case study even though I am fully aware that the circumstances are not real.
    \item Knowing that the patient and circumstances are made up makes it hard to feel real empathy.
    \item Being unable to feel real empathy for the patient makes it much harder to express empathy to the patient.
    \item Regardless of feeling real empathy, I believe that the appearance of being empathetic can be taught and learned.
    \item I felt like my “Empathize” communication skills improved between my first and second interaction with the Human Standardized Patient.
    \item Overall, I would give myself a (1-5) rating on “Empathize.”
    \item I felt like my “be Explicit” communication skills improved between my first and second interaction with the Human Standardized patient.
    \item Overall, I would give myself a (1-5) rating on “be Explicit.”
    \item I felt like my “Empower” communication skills improved between my first and second interaction with the Human Standardized patient.
    \item Overall, I would give myself a (1-5) rating on “Empower.”
    \item I felt like my “Overall” communication skills improved between my first and second interaction with the Human Standardized Patient.
    \item Overall, I would give myself a (1-5) rating on “Overall Communication Skills.”
    \item How likely would you do the procedure again to improve or refresh communication skills?
\end{enumerate} \\
\hline
\textbf{Stress-Related Questions} \\
\hline
Please rate the following statements (1-5) with 5 representing strong agreement:
\begin{enumerate}
    \item I was stressed during my interactions with the human standardized patients.
    \item It is easier for me to learn when I am relaxed and not experiencing stress.
\end{enumerate} \\
\hline
\end{tabular}
\end{table}

\begin{table}[htbp]
\centering
\caption{Arm-Specific Survey Questions}
\label{tab:arm_specific_questions}
\begin{tabular}{p{14cm}}
\hline
\textbf{Control Arm} \\
\hline
\begin{enumerate}
    \item I felt engaged during my reading of the paper.
    \item I felt that reading the paper was useful for learning communication skills.
    \item I anticipate using skills learned from doing the control procedure in clinical scenarios.
    \item I would be willing to read the SAME paper again in the future to improve or refresh my communication skills.
    \item I believe I would improve with this repeated re-reading.
\end{enumerate} \\
\hline
\textbf{SOPHIE Arm} \\
\hline
\begin{enumerate}
    \item I felt engaged during the SOPHIE modules.
    \item I felt that the SOPHIE modules were useful for learning communication skills.
    \item I anticipate using skills learned from the SOPHIE modules in clinical scenarios.
    \item I would be willing to engage with the SAME SOPHIE modules again in the future to improve or refresh my communication skills.
    \item I would be willing to engage with DIFFERENT SOPHIE modules again in the future to improve or refresh my communication skills.
    \item Receiving incentives (e.g., CME credits) would significantly increase my willingness to use SOPHIE.
    \item I was stressed while completing the SOPHIE modules.
    \item Interacting with an AI is less stressful than interacting with a real human.
    \item Getting personalized feedback on communication skill performance from a computer would help me improve.
    \item Having the ability to practice communication skills is essential for improving them.
\end{enumerate} \\
\hline
\end{tabular}
\end{table}

\begin{table}[htbp]
\centering
\caption{Shared interview questions asked of both Control and SOPHIE participants. The interviews explored participant perceptions of the training experience, communication skills learned, and future application in clinical practice.}
\label{tab:interview_shared}
\begin{tabular}{p{14cm}}
\hline
\textbf{Shared Interview Questions} \\
\hline
\begin{enumerate}
    \item What did you learn from participating in the experiment?
    \item What do you think were the most important messages?
    \item How did the program you did make you feel?
    \item What did you like most about doing the program?
    \item What do you think you will take away from the training into your future clinical practices?
    \item Please tell me everything you can recall about the three E’s.
    \item Did your training (reading a paper vs SOPHIE) help you feel more competent and/or confident Empathizing, being Explicit and Empowering? Why?
    \item If you were to be a user of a computer program designed to train communication skills, what would that experience look like?
    \item Would you prefer an interactive system, passive viewing materials, or a combination for end-of-life communication skills training? Why?
    \item Any questions, concerns, or additional thoughts you would like to share with the research team?
\end{enumerate} \\
\hline
\end{tabular}
\end{table}

\begin{table}[h]
    \centering
    \caption{SOPHIE UI/UX Survey: System Interaction and Perception}
    \label{tab:UIUX_survey1}
    \begin{tabular}{p{15cm}}
        \toprule
        \textbf{System Usability} \\
        \midrule
        \begin{enumerate}
            \item I found SOPHIE system easy to use.
            \item I think that I would like to use this system frequently.
            \item I do NOT think that I would need the support of a technical person to be able to use this system.
            \item I did NOT need to learn a lot of things before I could get going with this system.
        \end{enumerate} \\
        \midrule
        \textbf{Avatar and Emotion Perception} \\
        \midrule
        \begin{enumerate}
            \item SOPHIE’s voice sounded human-like.
            \item SOPHIE's voice showed emotion.
            \item SOPHIE's voice showed different emotions.
            \item SOPHIE's voice expressed sadness at times.
            \item SOPHIE's voice expressed fear or concern at times.
            \item SOPHIE's voice expressed frustration at times.
            \item SOPHIE looked realistic.
            \item SOPHIE's facial expressions expressed emotion.
            \item SOPHIE's facial expressions showed different emotions.
            \item SOPHIE's facial expressions expressed sadness at times.
            \item SOPHIE's facial expressions expressed concern at times.
            \item SOPHIE's facial expressions expressed frustration at times.
            \item SOPHIE, overall, seemed to be able to express emotions through face and voice.
            \item SOPHIE expressing emotion through face and voice simultaneously made the emotion easier to recognize.
            \item SOPHIE expressing emotion through face and voice simultaneously made the emotion more realistic and human-like.
        \end{enumerate} \\
        \midrule
        \textbf{Dialogue} \\
        \midrule
        \begin{enumerate}
            \item I found SOPHIE’s responses to be fluent and natural.
            \item SOPHIE’s questions/comments were relevant to her medical condition.
            \item SOPHIE’s responses followed logically from the conversation.
            \item SOPHIE reminded me of a real patient.
            \item I felt that SOPHIE understood what I said.
            \item The words SOPHIE used expressed her emotions.
        \end{enumerate} \\
        \bottomrule
    \end{tabular}
\end{table}

\begin{table}[h]
    \centering
    \caption{SOPHIE UI/UX Survey: Feedback Format and Skill Modules}
    \label{tab:UIUX_survey2}
    \begin{tabular}{p{15cm}}
        \toprule
        \textbf{Feedback Format} \\
        \midrule
        \begin{enumerate}
            \item I liked the overall format/structure in which the feedback was presented.
            \item I liked the way in which the homepage (e.g., first page seen) was presented.
            \item I liked the way in which the full feedback (e.g., all feedback when view full feedback button is clicked) was presented.
        \end{enumerate} \\
        \midrule
        \textbf{Skill-Specific Feedback (repeated across 3 modules)} \\
        \midrule
        \textit{Each of the following applies to Empathize, be Explicit, and Empower modules:}
        \begin{enumerate}
            \item I found the “Here is what you did well" section useful.
            \item I found the "here are more opportunities" section useful.
            \item I found the "View Suggestion" (e.g., system recommendations on homepage) section useful.
            \item I found the "full transcript with skills highlighted" section useful.
        \end{enumerate} \\
        \midrule
        \textbf{Additional Module-Specific Feedback} \\
        \midrule
        \textit{be Explicit Module:}
        \begin{enumerate}
            \item I found the "Hedge Words" feedback useful.
            \item I found the "Speaking Rate" feedback useful.
            \item I found the "Reading Level" feedback useful.
            \item Overall, I thought that the way the system quantified "be Explicit" was satisfactory.
        \end{enumerate}
        \textit{Empower Module:}
        \begin{enumerate}
            \item I found the "Questions Asked" feedback useful.
            \item I found the "Open-Ended Questions Asked" useful.
            \item I found the "Turn Taking" feedback useful.
            \item SOPHIE's responses gave me an opportunity to be empowering.
            \item Overall, I thought that the way the system quantified "Empower" was satisfactory.
        \end{enumerate} \\
        \bottomrule
    \end{tabular}
\end{table}

\begin{figure}[h]
    \centering
    \includegraphics[width=0.7\linewidth]{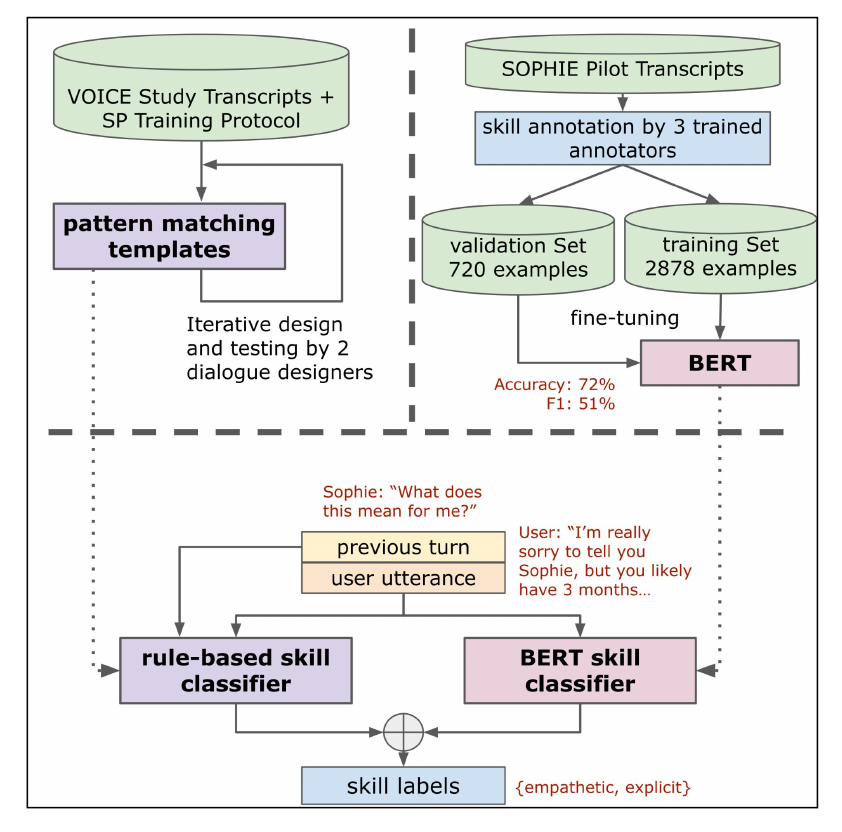}
    \caption{A hybrid skill classifier that detects whether a statement from the user falls under the 3E skills .}
    \label{fig:skill-classifier}
\end{figure}

\begin{figure}
    \centering
    \includegraphics[width=0.9\linewidth]{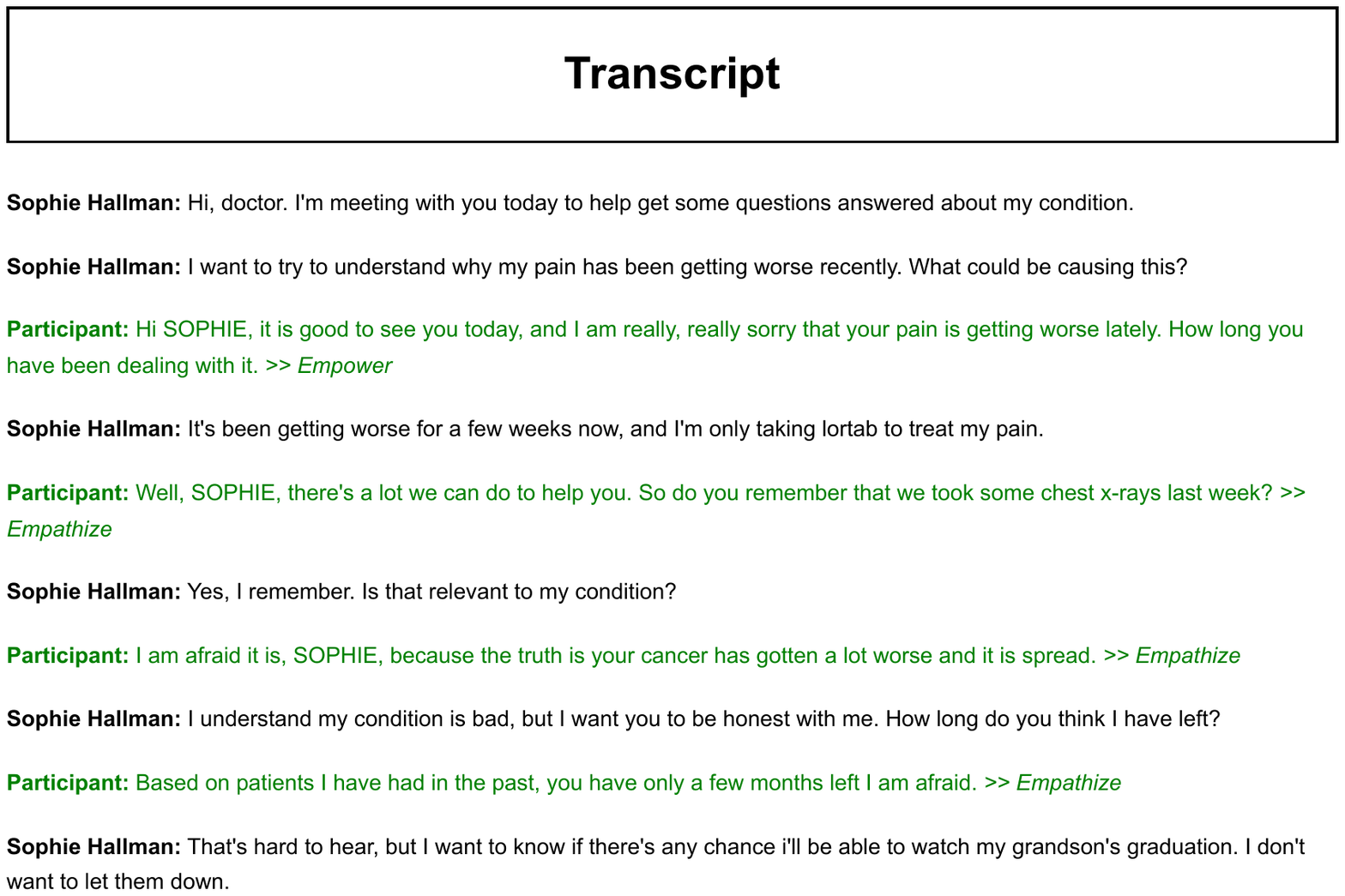}
    \caption{A part of a sample transcript from the Detailed Feedback with the 3E skill classification labeled by the classifier in Figure \ref{fig:skill-classifier}.}
    \label{fig:transcript}
\end{figure}

\begin{figure}
    \centering
    \includegraphics[width=\linewidth]{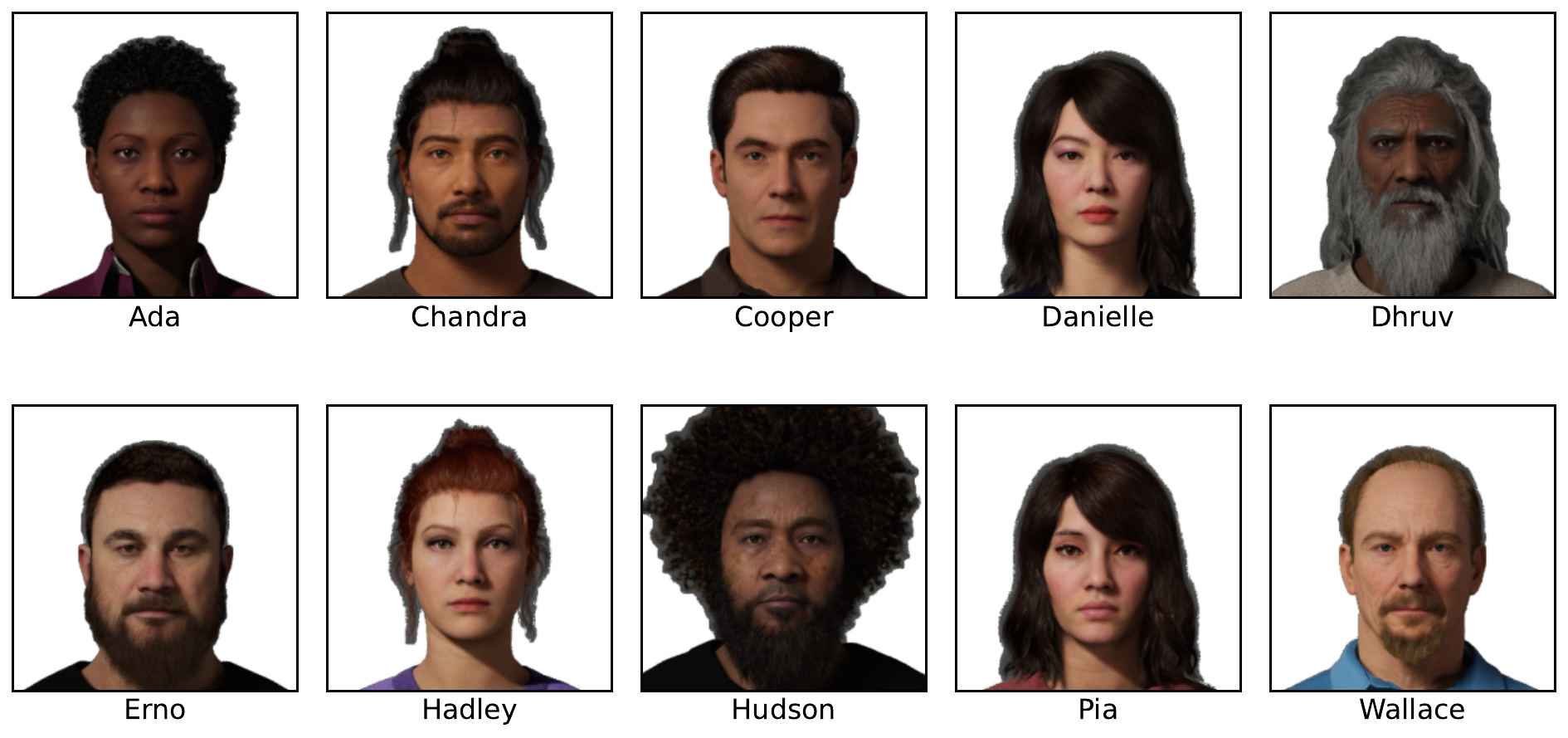}
    \caption{Virtual avatars can be customized to cover a Wide range of demographics including race, gender, age and cultural background. This could potentiality Enable clinicians to practice highly sensitive communication in ways that were previously difficult to achieve at scale with real humans.}
    \label{fig:different_races, genders, ages}
\end{figure}


%% file: main.bbl

\begin{thebibliography}{58}
\ifx \bisbn   \undefined \def \bisbn  #1{ISBN #1}\fi
\ifx \binits  \undefined \def \binits#1{#1}\fi
\ifx \bauthor  \undefined \def \bauthor#1{#1}\fi
\ifx \batitle  \undefined \def \batitle#1{#1}\fi
\ifx \bjtitle  \undefined \def \bjtitle#1{#1}\fi
\ifx \bvolume  \undefined \def \bvolume#1{\textbf{#1}}\fi
\ifx \byear  \undefined \def \byear#1{#1}\fi
\ifx \bissue  \undefined \def \bissue#1{#1}\fi
\ifx \bfpage  \undefined \def \bfpage#1{#1}\fi
\ifx \blpage  \undefined \def \blpage #1{#1}\fi
\ifx \burl  \undefined \def \burl#1{\textsf{#1}}\fi
\ifx \doiurl  \undefined \def \doiurl#1{\url{https://doi.org/#1}}\fi
\ifx \betal  \undefined \def \betal{\textit{et al.}}\fi
\ifx \binstitute  \undefined \def \binstitute#1{#1}\fi
\ifx \binstitutionaled  \undefined \def \binstitutionaled#1{#1}\fi
\ifx \bctitle  \undefined \def \bctitle#1{#1}\fi
\ifx \beditor  \undefined \def \beditor#1{#1}\fi
\ifx \bpublisher  \undefined \def \bpublisher#1{#1}\fi
\ifx \bbtitle  \undefined \def \bbtitle#1{#1}\fi
\ifx \bedition  \undefined \def \bedition#1{#1}\fi
\ifx \bseriesno  \undefined \def \bseriesno#1{#1}\fi
\ifx \blocation  \undefined \def \blocation#1{#1}\fi
\ifx \bsertitle  \undefined \def \bsertitle#1{#1}\fi
\ifx \bsnm \undefined \def \bsnm#1{#1}\fi
\ifx \bsuffix \undefined \def \bsuffix#1{#1}\fi
\ifx \bparticle \undefined \def \bparticle#1{#1}\fi
\ifx \barticle \undefined \def \barticle#1{#1}\fi
\bibcommenthead
\ifx \bconfdate \undefined \def \bconfdate #1{#1}\fi
\ifx \botherref \undefined \def \botherref #1{#1}\fi
\ifx \url \undefined \def \url#1{\textsf{#1}}\fi
\ifx \bchapter \undefined \def \bchapter#1{#1}\fi
\ifx \bbook \undefined \def \bbook#1{#1}\fi
\ifx \bcomment \undefined \def \bcomment#1{#1}\fi
\ifx \oauthor \undefined \def \oauthor#1{#1}\fi
\ifx \citeauthoryear \undefined \def \citeauthoryear#1{#1}\fi
\ifx \endbibitem  \undefined \def \endbibitem {}\fi
\ifx \bconflocation  \undefined \def \bconflocation#1{#1}\fi
\ifx \arxivurl  \undefined \def \arxivurl#1{\textsf{#1}}\fi
\csname PreBibitemsHook\endcsname

\bibitem[\protect\citeauthoryear{Shilling et~al.}{2024}]{shilling2024let}
\begin{barticle}
\bauthor{\bsnm{Shilling}, \binits{D.M.}},
\bauthor{\bsnm{Manz}, \binits{C.R.}},
\bauthor{\bsnm{Strand}, \binits{J.J.}},
\bauthor{\bsnm{Patel}, \binits{M.I.}}:
\batitle{Let us have the conversation: serious illness communication in oncology: definitions, barriers, and successful approaches}.
\bjtitle{American Society of Clinical Oncology Educational Book}
\bvolume{44}(\bissue{3}),
\bfpage{431352}
(\byear{2024})
\end{barticle}
\endbibitem

\bibitem[\protect\citeauthoryear{Pasricha et~al.}{2020}]{pasricha2020use}
\begin{barticle}
\bauthor{\bsnm{Pasricha}, \binits{V.}},
\bauthor{\bsnm{Gorman}, \binits{D.}},
\bauthor{\bsnm{Laothamatas}, \binits{K.}},
\bauthor{\bsnm{Bhardwaj}, \binits{A.}},
\bauthor{\bsnm{Ganta}, \binits{N.}},
\bauthor{\bsnm{Mikkelsen}, \binits{M.E.}}:
\batitle{Use of the serious illness conversation guide to improve communication with surrogates of critically ill patients. a pilot study}.
\bjtitle{ATS scholar}
\bvolume{1}(\bissue{2}),
\bfpage{119}--\blpage{133}
(\byear{2020})
\end{barticle}
\endbibitem

\bibitem[\protect\citeauthoryear{Paladino et~al.}{2023}]{paladino2023improving}
\begin{barticle}
\bauthor{\bsnm{Paladino}, \binits{J.}},
\bauthor{\bsnm{Sanders}, \binits{J.J.}},
\bauthor{\bsnm{Fromme}, \binits{E.K.}},
\bauthor{\bsnm{Block}, \binits{S.}},
\bauthor{\bsnm{Jacobsen}, \binits{J.C.}},
\bauthor{\bsnm{Jackson}, \binits{V.A.}},
\bauthor{\bsnm{Ritchie}, \binits{C.S.}},
\bauthor{\bsnm{Mitchell}, \binits{S.}}:
\batitle{Improving serious illness communication: a qualitative study of clinical culture}.
\bjtitle{BMC palliative care}
\bvolume{22}(\bissue{1}),
\bfpage{104}
(\byear{2023})
\end{barticle}
\endbibitem

\bibitem[\protect\citeauthoryear{Myers et~al.}{2024}]{myers2024simplifying}
\begin{barticle}
\bauthor{\bsnm{Myers}, \binits{J.}},
\bauthor{\bsnm{Steinberg}, \binits{L.}},
\bauthor{\bsnm{Incardona}, \binits{N.}},
\bauthor{\bsnm{Simon}, \binits{J.}},
\bauthor{\bsnm{Sanders}, \binits{J.}},
\bauthor{\bsnm{Seow}, \binits{H.}}:
\batitle{Simplifying serious illness communication: Preparing or deciding}.
\bjtitle{Current Oncology}
\bvolume{31}(\bissue{10}),
\bfpage{5832}--\blpage{5837}
(\byear{2024})
\end{barticle}
\endbibitem

\bibitem[\protect\citeauthoryear{Hagerty et~al.}{2005}]{hagerty2005communicating}
\begin{barticle}
\bauthor{\bsnm{Hagerty}, \binits{R.}},
\bauthor{\bsnm{Butow}, \binits{P.N.}},
\bauthor{\bsnm{Ellis}, \binits{P.}},
\bauthor{\bsnm{Dimitry}, \binits{S.}},
\bauthor{\bsnm{Tattersall}, \binits{M.}}:
\batitle{Communicating prognosis in cancer care: a systematic review of the literature}.
\bjtitle{Annals of oncology}
\bvolume{16}(\bissue{7}),
\bfpage{1005}--\blpage{1053}
(\byear{2005})
\end{barticle}
\endbibitem

\bibitem[\protect\citeauthoryear{Korsch and Negrete}{1972}]{korsch1972doctor}
\begin{barticle}
\bauthor{\bsnm{Korsch}, \binits{B.M.}},
\bauthor{\bsnm{Negrete}, \binits{V.F.}}:
\batitle{Doctor-patient communication}.
\bjtitle{Scientific American}
\bvolume{227}(\bissue{2}),
\bfpage{66}--\blpage{75}
(\byear{1972})
\end{barticle}
\endbibitem

\bibitem[\protect\citeauthoryear{Ha and Longnecker}{2010}]{ha2010doctor}
\begin{barticle}
\bauthor{\bsnm{Ha}, \binits{J.F.}},
\bauthor{\bsnm{Longnecker}, \binits{N.}}:
\batitle{Doctor-patient communication: a review}.
\bjtitle{Ochsner journal}
\bvolume{10}(\bissue{1}),
\bfpage{38}--\blpage{43}
(\byear{2010})
\end{barticle}
\endbibitem

\bibitem[\protect\citeauthoryear{Riedl and Sch{\"u}{\ss}ler}{2017}]{riedl2017influence}
\begin{barticle}
\bauthor{\bsnm{Riedl}, \binits{D.}},
\bauthor{\bsnm{Sch{\"u}{\ss}ler}, \binits{G.}}:
\batitle{The influence of doctor-patient communication on health outcomes: a systematic review}.
\bjtitle{Zeitschrift f{\"u}r Psychosomatische Medizin und Psychotherapie}
\bvolume{63}(\bissue{2}),
\bfpage{131}--\blpage{150}
(\byear{2017})
\end{barticle}
\endbibitem

\bibitem[\protect\citeauthoryear{Stewart}{1995}]{stewart1995effective}
\begin{barticle}
\bauthor{\bsnm{Stewart}, \binits{M.A.}}:
\batitle{Effective physician-patient communication and health outcomes: a review.}
\bjtitle{CMAJ: Canadian medical association journal}
\bvolume{152}(\bissue{9}),
\bfpage{1423}
(\byear{1995})
\end{barticle}
\endbibitem

\bibitem[\protect\citeauthoryear{Begum}{2014}]{begum2014doctor}
\begin{barticle}
\bauthor{\bsnm{Begum}, \binits{T.}}:
\batitle{Doctor patient communication: a review}.
\bjtitle{Journal of Bangladesh College of Physicians and Surgeons}
\bvolume{32}(\bissue{2}),
\bfpage{84}--\blpage{88}
(\byear{2014})
\end{barticle}
\endbibitem

\bibitem[\protect\citeauthoryear{for Disease~Control et~al.}{2015}]{centers2015national}
\begin{botherref}
\oauthor{\bsnm{Disease~Control}, \binits{C.}},
\oauthor{\bsnm{Prevention}}, et al.:
National ambulatory medical care survey: 2012 state and national summary tables.
Washington: Centers for Disease Control and Prevention
(2015)
\end{botherref}
\endbibitem

\bibitem[\protect\citeauthoryear{Oates et~al.}{2000}]{oates2000impact}
\begin{barticle}
\bauthor{\bsnm{Oates}, \binits{J.}},
\bauthor{\bsnm{Weston}, \binits{W.W.}},
\bauthor{\bsnm{Jordan}, \binits{J.}}, \betal:
\batitle{The impact of patient-centered care on outcomes}.
\bjtitle{Fam Pract}
\bvolume{49}(\bissue{9}),
\bfpage{796}--\blpage{804}
(\byear{2000})
\end{barticle}
\endbibitem

\bibitem[\protect\citeauthoryear{Beck et~al.}{2002}]{beck2002physician}
\begin{barticle}
\bauthor{\bsnm{Beck}, \binits{R.S.}},
\bauthor{\bsnm{Daughtridge}, \binits{R.}},
\bauthor{\bsnm{Sloane}, \binits{P.D.}}:
\batitle{Physician-patient communication in the primary care office: a systematic review.}
\bjtitle{The Journal of the American Board of Family Practice}
\bvolume{15}(\bissue{1}),
\bfpage{25}--\blpage{38}
(\byear{2002})
\end{barticle}
\endbibitem

\bibitem[\protect\citeauthoryear{Back and Curtis}{2002}]{back2002communicating}
\begin{barticle}
\bauthor{\bsnm{Back}, \binits{A.L.}},
\bauthor{\bsnm{Curtis}, \binits{J.R.}}:
\batitle{Communicating bad news}.
\bjtitle{Western Journal of Medicine}
\bvolume{176}(\bissue{3}),
\bfpage{177}
(\byear{2002})
\end{barticle}
\endbibitem

\bibitem[\protect\citeauthoryear{Gessesse et~al.}{2023}]{gessesse2023exploring}
\begin{botherref}
\oauthor{\bsnm{Gessesse}, \binits{A.G.}},
\oauthor{\bsnm{Haile}, \binits{J.M.}},
\oauthor{\bsnm{Woldearegay}, \binits{A.G.}}:
Exploring effective communication strategies employed by physicians in delivering bad news in ethiopian state hospitals.
Patient Related Outcome Measures,
409--425
(2023)
\end{botherref}
\endbibitem

\bibitem[\protect\citeauthoryear{Bagley}{2023}]{bagley2023delivering}
\begin{botherref}
\oauthor{\bsnm{Bagley}, \binits{J.}}:
Delivering difficult patient conversations is a skill to be learned, practiced.
Bulletin of The American College of Surgeons,
8--13
(2023)
\end{botherref}
\endbibitem

\bibitem[\protect\citeauthoryear{Starcke and Brand}{2012}]{starcke2012decision}
\begin{barticle}
\bauthor{\bsnm{Starcke}, \binits{K.}},
\bauthor{\bsnm{Brand}, \binits{M.}}:
\batitle{Decision making under stress: a selective review}.
\bjtitle{Neuroscience \& Biobehavioral Reviews}
\bvolume{36}(\bissue{4}),
\bfpage{1228}--\blpage{1248}
(\byear{2012})
\end{barticle}
\endbibitem

\bibitem[\protect\citeauthoryear{Arnsten}{2009}]{arnsten2009stress}
\begin{barticle}
\bauthor{\bsnm{Arnsten}, \binits{A.F.}}:
\batitle{Stress signalling pathways that impair prefrontal cortex structure and function}.
\bjtitle{Nature reviews neuroscience}
\bvolume{10}(\bissue{6}),
\bfpage{410}--\blpage{422}
(\byear{2009})
\end{barticle}
\endbibitem

\bibitem[\protect\citeauthoryear{Gruneir et~al.}{2007}]{gruneir2007people}
\begin{barticle}
\bauthor{\bsnm{Gruneir}, \binits{A.}},
\bauthor{\bsnm{Mor}, \binits{V.}},
\bauthor{\bsnm{Weitzen}, \binits{S.}},
\bauthor{\bsnm{Truchil}, \binits{R.}},
\bauthor{\bsnm{Teno}, \binits{J.}},
\bauthor{\bsnm{Roy}, \binits{J.}}:
\batitle{Where people die: a multilevel approach to understanding influences on site of death in america}.
\bjtitle{Medical Care Research and Review}
\bvolume{64}(\bissue{4}),
\bfpage{351}--\blpage{378}
(\byear{2007})
\end{barticle}
\endbibitem

\bibitem[\protect\citeauthoryear{Humphrey et~al.}{2022}]{humphrey2022frequency}
\begin{barticle}
\bauthor{\bsnm{Humphrey}, \binits{K.E.}},
\bauthor{\bsnm{Sundberg}, \binits{M.}},
\bauthor{\bsnm{Milliren}, \binits{C.E.}},
\bauthor{\bsnm{Graham}, \binits{D.A.}},
\bauthor{\bsnm{Landrigan}, \binits{C.P.}}:
\batitle{Frequency and nature of communication and handoff failures in medical malpractice claims}.
\bjtitle{Journal of patient safety}
\bvolume{18}(\bissue{2}),
\bfpage{130}--\blpage{137}
(\byear{2022})
\end{barticle}
\endbibitem

\bibitem[\protect\citeauthoryear{Epstein and r.L Street}{2007}]{healingpatient}
\begin{botherref}
\oauthor{\bsnm{Epstein}, \binits{R.M.}},
\oauthor{\bsnm{Street}, \binits{J.}}:
Patient-centered communication in cancer care.
National Cancer Institute
(2007)
\end{botherref}
\endbibitem

\bibitem[\protect\citeauthoryear{Gilligan et~al.}{2017}]{gilligan2017patient}
\begin{barticle}
\bauthor{\bsnm{Gilligan}, \binits{T.}},
\bauthor{\bsnm{Coyle}, \binits{N.}},
\bauthor{\bsnm{Frankel}, \binits{R.M.}},
\bauthor{\bsnm{Berry}, \binits{D.L.}},
\bauthor{\bsnm{Bohlke}, \binits{K.}},
\bauthor{\bsnm{Epstein}, \binits{R.M.}},
\bauthor{\bsnm{Finlay}, \binits{E.}},
\bauthor{\bsnm{Jackson}, \binits{V.A.}},
\bauthor{\bsnm{Lathan}, \binits{C.S.}},
\bauthor{\bsnm{Loprinzi}, \binits{C.L.}}, \betal:
\batitle{Patient-clinician communication: American society of clinical oncology consensus guideline}.
\bjtitle{Journal of Clinical Oncology}
\bvolume{35}(\bissue{31}),
\bfpage{3618}--\blpage{3632}
(\byear{2017})
\end{barticle}
\endbibitem

\bibitem[\protect\citeauthoryear{Bosse et~al.}{2015}]{bosse2015cost}
\begin{barticle}
\bauthor{\bsnm{Bosse}, \binits{H.M.}},
\bauthor{\bsnm{Nickel}, \binits{M.}},
\bauthor{\bsnm{Huwendiek}, \binits{S.}},
\bauthor{\bsnm{Schultz}, \binits{J.H.}},
\bauthor{\bsnm{Nikendei}, \binits{C.}}:
\batitle{Cost-effectiveness of peer role play and standardized patients in undergraduate communication training}.
\bjtitle{BMC medical education}
\bvolume{15},
\bfpage{1}--\blpage{6}
(\byear{2015})
\end{barticle}
\endbibitem

\bibitem[\protect\citeauthoryear{Gillette et~al.}{2017}]{gillette2017cost}
\begin{barticle}
\bauthor{\bsnm{Gillette}, \binits{C.}},
\bauthor{\bsnm{Stanton}, \binits{R.B.}},
\bauthor{\bsnm{Rockich-Winston}, \binits{N.}},
\bauthor{\bsnm{Rudolph}, \binits{M.}},
\bauthor{\bsnm{Anderson~Jr}, \binits{H.G.}}:
\batitle{Cost-effectiveness of using standardized patients to assess student-pharmacist communication skills}.
\bjtitle{American journal of pharmaceutical education}
\bvolume{81}(\bissue{10}),
\bfpage{6120}
(\byear{2017})
\end{barticle}
\endbibitem

\bibitem[\protect\citeauthoryear{Menez et~al.}{2024}]{menez2024strategies}
\begin{botherref}
\oauthor{\bsnm{Menez}, \binits{O.}},
\oauthor{\bsnm{Zagales}, \binits{R.}},
\oauthor{\bsnm{Colleton}, \binits{T.}},
\oauthor{\bsnm{Rodriguez~Artze}, \binits{C.}},
\oauthor{\bsnm{Cordero}, \binits{L.}}:
Strategies and challenges to diversifying standardized patients at a rural regional campus
(2024)
\end{botherref}
\endbibitem

\bibitem[\protect\citeauthoryear{Everett et~al.}{2005}]{everett2005recruitment}
\begin{barticle}
\bauthor{\bsnm{Everett}, \binits{M.}},
\bauthor{\bsnm{May}, \binits{W.}},
\bauthor{\bsnm{Nowels}, \binits{C.}},
\bauthor{\bsnm{Main}, \binits{D.}}:
\batitle{Recruitment, retention, and training of african american and latino standardized patients: a collaborative study}.
\bjtitle{Med Sci Educ}
\bvolume{15}(\bissue{2}),
\bfpage{74}--\blpage{80}
(\byear{2005})
\end{barticle}
\endbibitem

\bibitem[\protect\citeauthoryear{Bowers et~al.}{2024}]{bowers2024artificial}
\begin{botherref}
\oauthor{\bsnm{Bowers}, \binits{P.}},
\oauthor{\bsnm{Graydon}, \binits{K.}},
\oauthor{\bsnm{Ryan}, \binits{T.}},
\oauthor{\bsnm{Lau}, \binits{J.H.}},
\oauthor{\bsnm{Tomlin}, \binits{D.}}:
Artificial intelligence-driven virtual patients for communication skill development in healthcare students: A scoping review.
Australasian Journal of Educational Technology
(2024)
\end{botherref}
\endbibitem

\bibitem[\protect\citeauthoryear{Wang et~al.}{2024}]{wang2024commsense}
\begin{barticle}
\bauthor{\bsnm{Wang}, \binits{Z.}},
\bauthor{\bsnm{Hassan}, \binits{N.}},
\bauthor{\bsnm{LeBaron}, \binits{V.}},
\bauthor{\bsnm{Flickinger}, \binits{T.}},
\bauthor{\bsnm{Ling}, \binits{D.}},
\bauthor{\bsnm{Edwards}, \binits{J.}},
\bauthor{\bsnm{Wu}, \binits{C.}},
\bauthor{\bsnm{Boukhechba}, \binits{M.}},
\bauthor{\bsnm{Barnes}, \binits{L.E.}}:
\batitle{Commsense: A wearable sensing computational framework for evaluating patient-clinician interactions}.
\bjtitle{Proceedings of the ACM on Human-Computer Interaction}
\bvolume{8}(\bissue{CSCW2}),
\bfpage{1}--\blpage{31}
(\byear{2024})
\end{barticle}
\endbibitem

\bibitem[\protect\citeauthoryear{Kearns et~al.}{2024}]{kearns2024bridging}
\begin{bchapter}
\bauthor{\bsnm{Kearns}, \binits{W.R.}},
\bauthor{\bsnm{Bertram}, \binits{J.}},
\bauthor{\bsnm{Divina}, \binits{M.}},
\bauthor{\bsnm{Kemp}, \binits{L.}},
\bauthor{\bsnm{Wang}, \binits{Y.}},
\bauthor{\bsnm{Marin}, \binits{A.}},
\bauthor{\bsnm{Cohen}, \binits{T.}},
\bauthor{\bsnm{Yuwen}, \binits{W.}}:
\bctitle{Bridging the skills gap: Evaluating an ai-assisted provider platform to support care providers with empathetic delivery of protocolized therapy}.
In: \bbtitle{AMIA Annual Symposium Proceedings},
vol. \bseriesno{2023},
p. \bfpage{436}
(\byear{2024})
\end{bchapter}
\endbibitem

\bibitem[\protect\citeauthoryear{Back and Arnold}{2009}]{back2009mastering}
\begin{bbook}
\bauthor{\bsnm{Back}, \binits{A.}},
\bauthor{\bsnm{Arnold}, \binits{R.}}:
\bbtitle{Mastering Communication with Seriously Ill Patients: Balancing Honesty with Empathy and Hope}.
\bpublisher{Cambridge University Press}, \blocation{???}
(\byear{2009})
\end{bbook}
\endbibitem

\bibitem[\protect\citeauthoryear{Elias et~al.}{2017}]{elias2017social}
\begin{barticle}
\bauthor{\bsnm{Elias}, \binits{C.M.}},
\bauthor{\bsnm{Shields}, \binits{C.G.}},
\bauthor{\bsnm{Griggs}, \binits{J.J.}},
\bauthor{\bsnm{Fiscella}, \binits{K.}},
\bauthor{\bsnm{Christ}, \binits{S.L.}},
\bauthor{\bsnm{Colbert}, \binits{J.}},
\bauthor{\bsnm{Henry}, \binits{S.G.}},
\bauthor{\bsnm{Hoh}, \binits{B.G.}},
\bauthor{\bsnm{Hunte}, \binits{H.E.}},
\bauthor{\bsnm{Marshall}, \binits{M.}}, \betal:
\batitle{The social and behavioral influences (sbi) study: study design and rationale for studying the effects of race and activation on cancer pain management}.
\bjtitle{BMC cancer}
\bvolume{17},
\bfpage{1}--\blpage{11}
(\byear{2017})
\end{barticle}
\endbibitem

\bibitem[\protect\citeauthoryear{Epstein et~al.}{2017}]{epstein2017effect}
\begin{barticle}
\bauthor{\bsnm{Epstein}, \binits{R.M.}},
\bauthor{\bsnm{Duberstein}, \binits{P.R.}},
\bauthor{\bsnm{Fenton}, \binits{J.J.}},
\bauthor{\bsnm{Fiscella}, \binits{K.}},
\bauthor{\bsnm{Hoerger}, \binits{M.}},
\bauthor{\bsnm{Tancredi}, \binits{D.J.}},
\bauthor{\bsnm{Xing}, \binits{G.}},
\bauthor{\bsnm{Gramling}, \binits{R.}},
\bauthor{\bsnm{Mohile}, \binits{S.}},
\bauthor{\bsnm{Franks}, \binits{P.}}, \betal:
\batitle{Effect of a patient-centered communication intervention on oncologist-patient communication, quality of life, and health care utilization in advanced cancer: the voice randomized clinical trial}.
\bjtitle{JAMA oncology}
\bvolume{3}(\bissue{1}),
\bfpage{92}--\blpage{100}
(\byear{2017})
\end{barticle}
\endbibitem

\bibitem[\protect\citeauthoryear{Epstein et~al.}{2001}]{epstein2001improving}
\begin{barticle}
\bauthor{\bsnm{Epstein}, \binits{R.M.}},
\bauthor{\bsnm{Levenkron}, \binits{J.C.}},
\bauthor{\bsnm{Frarey}, \binits{L.}},
\bauthor{\bsnm{Thompson}, \binits{J.}},
\bauthor{\bsnm{Anderson}, \binits{K.}},
\bauthor{\bsnm{Franks}, \binits{P.}}:
\batitle{Improving physicians’ hiv risk-assessment skills using announced and unannounced standardized patients}.
\bjtitle{Journal of general internal medicine}
\bvolume{16},
\bfpage{176}--\blpage{180}
(\byear{2001})
\end{barticle}
\endbibitem

\bibitem[\protect\citeauthoryear{Hoerger et~al.}{2013}]{hoerger2013values}
\begin{barticle}
\bauthor{\bsnm{Hoerger}, \binits{M.}},
\bauthor{\bsnm{Epstein}, \binits{R.M.}},
\bauthor{\bsnm{Winters}, \binits{P.C.}},
\bauthor{\bsnm{Fiscella}, \binits{K.}},
\bauthor{\bsnm{Duberstein}, \binits{P.R.}},
\bauthor{\bsnm{Gramling}, \binits{R.}},
\bauthor{\bsnm{Butow}, \binits{P.N.}},
\bauthor{\bsnm{Mohile}, \binits{S.G.}},
\bauthor{\bsnm{Kaesberg}, \binits{P.R.}},
\bauthor{\bsnm{Tang}, \binits{W.}}, \betal:
\batitle{Values and options in cancer care (voice): study design and rationale for a patient-centered communication and decision-making intervention for physicians, patients with advanced cancer, and their caregivers}.
\bjtitle{BMC cancer}
\bvolume{13},
\bfpage{1}--\blpage{14}
(\byear{2013})
\end{barticle}
\endbibitem

\bibitem[\protect\citeauthoryear{Horowitz et~al.}{2020}]{horowitz2020mvp}
\begin{barticle}
\bauthor{\bsnm{Horowitz}, \binits{R.K.}},
\bauthor{\bsnm{Hogan}, \binits{L.A.}},
\bauthor{\bsnm{Carroll}, \binits{T.}}:
\batitle{Mvp--medical situation, values, and plan: A memorable and useful model for all serious illness conversations}.
\bjtitle{Journal of pain and symptom management}
\bvolume{60}(\bissue{5}),
\bfpage{1059}--\blpage{1065}
(\byear{2020})
\end{barticle}
\endbibitem

\bibitem[\protect\citeauthoryear{Sen et~al.}{2017}]{sen2017modeling}
\begin{bchapter}
\bauthor{\bsnm{Sen}, \binits{T.}},
\bauthor{\bsnm{Ali}, \binits{M.R.}},
\bauthor{\bsnm{Hoque}, \binits{M.E.}},
\bauthor{\bsnm{Epstein}, \binits{R.}},
\bauthor{\bsnm{Duberstein}, \binits{P.}}:
\bctitle{Modeling doctor-patient communication with affective text analysis}.
In: \bbtitle{2017 Seventh International Conference on Affective Computing and Intelligent Interaction (ACII)},
pp. \bfpage{170}--\blpage{177}
(\byear{2017}).
\bcomment{IEEE}
\end{bchapter}
\endbibitem

\bibitem[\protect\citeauthoryear{Kane et~al.}{2023}]{kane2023managing}
\begin{barticle}
\bauthor{\bsnm{Kane}, \binits{B.}},
\bauthor{\bsnm{Giugno}, \binits{C.}},
\bauthor{\bsnm{Schubert}, \binits{L.}},
\bauthor{\bsnm{Haut}, \binits{K.}},
\bauthor{\bsnm{Wohn}, \binits{C.}},
\bauthor{\bsnm{Hoque}, \binits{E.}}:
\batitle{Managing emotional dialogue for a virtual cancer patient: A schema-guided approach}.
\bjtitle{IEEE Transactions on Affective Computing}
\bvolume{15}(\bissue{3}),
\bfpage{1041}--\blpage{1052}
(\byear{2023})
\end{barticle}
\endbibitem

\bibitem[\protect\citeauthoryear{Ali et~al.}{2021}]{ali2021novel}
\begin{barticle}
\bauthor{\bsnm{Ali}, \binits{M.R.}},
\bauthor{\bsnm{Sen}, \binits{T.}},
\bauthor{\bsnm{Kane}, \binits{B.}},
\bauthor{\bsnm{Bose}, \binits{S.}},
\bauthor{\bsnm{Carroll}, \binits{T.M.}},
\bauthor{\bsnm{Epstein}, \binits{R.}},
\bauthor{\bsnm{Schubert}, \binits{L.}},
\bauthor{\bsnm{Hoque}, \binits{E.}}:
\batitle{Novel computational linguistic measures, dialogue system and the development of sophie: Standardized online patient for healthcare interaction education}.
\bjtitle{IEEE Transactions on Affective Computing}
\bvolume{14}(\bissue{1}),
\bfpage{223}--\blpage{235}
(\byear{2021})
\end{barticle}
\endbibitem

\bibitem[\protect\citeauthoryear{Haut et~al.}{2023}]{haut2023validating}
\begin{bchapter}
\bauthor{\bsnm{Haut}, \binits{K.}},
\bauthor{\bsnm{Wohn}, \binits{C.}},
\bauthor{\bsnm{Kane}, \binits{B.}},
\bauthor{\bsnm{Carroll}, \binits{T.}},
\bauthor{\bsnm{Guigno}, \binits{C.}},
\bauthor{\bsnm{Kumar}, \binits{V.}},
\bauthor{\bsnm{Epstein}, \binits{R.}},
\bauthor{\bsnm{Schuber}, \binits{L.}},
\bauthor{\bsnm{Hoque}, \binits{E.}}:
\bctitle{Validating a virtual human and automated feedback system for training doctor-patient communication skills}.
In: \bbtitle{2023 11th International Conference on Affective Computing and Intelligent Interaction (ACII)},
pp. \bfpage{1}--\blpage{8}
(\byear{2023}).
\bcomment{IEEE}
\end{bchapter}
\endbibitem

\bibitem[\protect\citeauthoryear{Bereiter and Scardamalia}{1993}]{bereiter1993surpassing}
\begin{botherref}
\oauthor{\bsnm{Bereiter}, \binits{C.}},
\oauthor{\bsnm{Scardamalia}, \binits{M.}}:
Surpassing ourselves.
An inquiry into the nature and implications of expertise. Chicago: Open Court
(1993)
\end{botherref}
\endbibitem

\bibitem[\protect\citeauthoryear{Ende}{1983}]{ende1983feedback}
\begin{barticle}
\bauthor{\bsnm{Ende}, \binits{J.}}:
\batitle{Feedback in clinical medical education}.
\bjtitle{Jama}
\bvolume{250}(\bissue{6}),
\bfpage{777}--\blpage{781}
(\byear{1983})
\end{barticle}
\endbibitem

\bibitem[\protect\citeauthoryear{Molloy et~al.}{2012}]{molloy2012impact}
\begin{bchapter}
\bauthor{\bsnm{Molloy}, \binits{E.}},
\bauthor{\bsnm{Borrell-Carrio}, \binits{F.}},
\bauthor{\bsnm{Epstein}, \binits{R.}}:
\bctitle{The impact of emotions in feedback}.
In: \bbtitle{Feedback in Higher and Professional Education},
pp. \bfpage{50}--\blpage{71}.
\bpublisher{Routledge}, \blocation{???}
(\byear{2012})
\end{bchapter}
\endbibitem

\bibitem[\protect\citeauthoryear{Boud and Molloy}{2013}]{boud2013feedback}
\begin{botherref}
\oauthor{\bsnm{Boud}, \binits{D.}},
\oauthor{\bsnm{Molloy}, \binits{E.}}:
Feedback in higher and professional education.
Understanding it and doing it well,
2013
(2013)
\end{botherref}
\endbibitem

\bibitem[\protect\citeauthoryear{Thomas and Arnold}{2011}]{thomas2011giving}
\begin{barticle}
\bauthor{\bsnm{Thomas}, \binits{J.D.}},
\bauthor{\bsnm{Arnold}, \binits{R.M.}}:
\batitle{Giving feedback}.
\bjtitle{Journal of palliative medicine}
\bvolume{14}(\bissue{2}),
\bfpage{233}--\blpage{239}
(\byear{2011})
\end{barticle}
\endbibitem

\bibitem[\protect\citeauthoryear{Horowitz et~al.}{2020}]{mvp}
\begin{barticle}
\bauthor{\bsnm{Horowitz}, \binits{R.K.}},
\bauthor{\bsnm{Hogan}, \binits{L.A.}},
\bauthor{\bsnm{Carroll}, \binits{T.}}:
\batitle{{MVP-Medical Situation, Values, and Plan: A Memorable and Useful Model for All Serious Illness Conversations}}.
\bjtitle{Journal of Pain and Symptom Management}
\bvolume{60}(\bissue{5}),
\bfpage{1059}--\blpage{1065}
(\byear{2020})
\doiurl{10.1016/j.jpainsymman.2020.07.022} .
\bcomment{Epub 2020 Jul 30}
\end{barticle}
\endbibitem

\bibitem[\protect\citeauthoryear{Devlin et~al.}{2019}]{bert}
\begin{bchapter}
\bauthor{\bsnm{Devlin}, \binits{J.}},
\bauthor{\bsnm{Chang}, \binits{M.-W.}},
\bauthor{\bsnm{Lee}, \binits{K.}},
\bauthor{\bsnm{Toutanova}, \binits{K.}}:
\bctitle{Bert: Pre-training of deep bidirectional transformers for language understanding}.
In: \bbtitle{Proceedings of the 2019 Conference of the North American Chapter of the Association for Computational Linguistics: Human Language Technologies, Volume 1 (long and Short Papers)},
pp. \bfpage{4171}--\blpage{4186}
(\byear{2019})
\end{bchapter}
\endbibitem

\bibitem[\protect\citeauthoryear{Zhu and Wu}{2025}]{chatbot1}
\begin{botherref}
\oauthor{\bsnm{Zhu}, \binits{J.}},
\oauthor{\bsnm{Wu}, \binits{J.}}:
Ask patients with patience: Enabling llms for human-centric medical dialogue with grounded reasoning.
arXiv preprint arXiv:2502.07143
(2025)
\end{botherref}
\endbibitem

\bibitem[\protect\citeauthoryear{Mehandru et~al.}{2024}]{chatbot2}
\begin{barticle}
\bauthor{\bsnm{Mehandru}, \binits{N.}},
\bauthor{\bsnm{Miao}, \binits{B.Y.}},
\bauthor{\bsnm{Almaraz}, \binits{E.R.}},
\bauthor{\bsnm{Sushil}, \binits{M.}},
\bauthor{\bsnm{Butte}, \binits{A.J.}},
\bauthor{\bsnm{Alaa}, \binits{A.}}:
\batitle{{Evaluating large language models as agents in the clinic}}.
\bjtitle{npj Digital Medicine}
\bvolume{7}(\bissue{1}),
\bfpage{84}
(\byear{2024})
\doiurl{10.1038/s41746-024-01083-y}
\end{barticle}
\endbibitem

\bibitem[\protect\citeauthoryear{Geantă et~al.}{2024}]{chatbot3}
\begin{barticle}
\bauthor{\bsnm{Geantă}, \binits{M.}},
\bauthor{\bsnm{Bădescu}, \binits{D.}},
\bauthor{\bsnm{Chirca}, \binits{N.}},
\bauthor{\bsnm{Nechita}, \binits{O.C.}},
\bauthor{\bsnm{Radu}, \binits{C.G.}},
\bauthor{\bsnm{Rascu}, \binits{S.}},
\bauthor{\bsnm{Rădăvoi}, \binits{D.}},
\bauthor{\bsnm{Sima}, \binits{C.}},
\bauthor{\bsnm{Toma}, \binits{C.}},
\bauthor{\bsnm{Jinga}, \binits{V.}}:
\batitle{{The Potential Impact of Large Language Models on Doctor-Patient Communication: A Case Study in Prostate Cancer}}.
\bjtitle{Healthcare (Basel)}
\bvolume{12}(\bissue{15}),
\bfpage{1548}
(\byear{2024})
\doiurl{10.3390/healthcare12151548} .
\bcomment{Epub 2024 Aug 5}
\end{barticle}
\endbibitem

\bibitem[\protect\citeauthoryear{Ali et~al.}{2015}]{lissa}
\begin{bchapter}
\bauthor{\bsnm{Ali}, \binits{M.R.}},
\bauthor{\bsnm{Crasta}, \binits{D.}},
\bauthor{\bsnm{Jin}, \binits{L.}},
\bauthor{\bsnm{Baretto}, \binits{A.}},
\bauthor{\bsnm{Pachter}, \binits{J.}},
\bauthor{\bsnm{Rogge}, \binits{R.D.}},
\bauthor{\bsnm{Hoque}, \binits{M.E.}}:
\bctitle{Lissa — live interactive social skill assistance}.
In: \bbtitle{2015 International Conference on Affective Computing and Intelligent Interaction (ACII)},
pp. \bfpage{173}--\blpage{179}
(\byear{2015}).
\doiurl{10.1109/ACII.2015.7344568}
\end{bchapter}
\endbibitem

\bibitem[\protect\citeauthoryear{Razavi et~al.}{2022}]{aging}
\begin{botherref}
\oauthor{\bsnm{Razavi}, \binits{S.Z.}},
\oauthor{\bsnm{Schubert}, \binits{L.K.}},
\oauthor{\bsnm{Orden}, \binits{K.}},
\oauthor{\bsnm{Ali}, \binits{M.R.}},
\oauthor{\bsnm{Kane}, \binits{B.}},
\oauthor{\bsnm{Hoque}, \binits{E.}}:
Discourse behavior of older adults interacting with a dialogue agent competent in multiple topics
\textbf{12}(2)
(2022)
\doiurl{10.1145/3484510}
\end{botherref}
\endbibitem

\bibitem[\protect\citeauthoryear{Engine}{}]{metahuman}
\begin{botherref}
\oauthor{\bsnm{Engine}, \binits{U.}}:
MetaHuman | Realistic Person Creator - Unreal Engine.
[Online; accessed 2025-03-16].
\url{https://www.unrealengine.com/en-US/metahuman}
\end{botherref}
\endbibitem

\bibitem[\protect\citeauthoryear{Radziej et~al.}{2017}]{radziej2017assess}
\begin{barticle}
\bauthor{\bsnm{Radziej}, \binits{K.}},
\bauthor{\bsnm{Loechner}, \binits{J.}},
\bauthor{\bsnm{Engerer}, \binits{C.}},
\bauthor{\bsnm{Figueiredo}, \binits{M.}},
\bauthor{\bsnm{Freund}, \binits{J.}},
\bauthor{\bsnm{Sattel}, \binits{H.}},
\bauthor{\bsnm{Bachmann}, \binits{C.}},
\bauthor{\bsnm{Berberat}, \binits{P.O.}},
\bauthor{\bsnm{Dinkel}, \binits{A.}},
\bauthor{\bsnm{Wuensch}, \binits{A.}}:
\batitle{How to assess communication skills? development of the rating scale comon check}.
\bjtitle{Medical education online}
\bvolume{22}(\bissue{1}),
\bfpage{1392823}
(\byear{2017})
\end{barticle}
\endbibitem

\bibitem[\protect\citeauthoryear{Chauhan et~al.}{2023}]{chauhan2023validated}
\begin{barticle}
\bauthor{\bsnm{Chauhan}, \binits{A.}},
\bauthor{\bsnm{Begum}, \binits{J.}},
\bauthor{\bsnm{Saiyad}, \binits{S.}}:
\batitle{Validated checklist for assessing communication skills in undergraduate medical students: bridging the gap for effective doctor-patient interactions}.
\bjtitle{Advances in Physiology Education}
\bvolume{47}(\bissue{4}),
\bfpage{871}--\blpage{879}
(\byear{2023})
\end{barticle}
\endbibitem

\bibitem[\protect\citeauthoryear{Stein et~al.}{2022}]{stein2022general}
\begin{barticle}
\bauthor{\bsnm{Stein}, \binits{D.}},
\bauthor{\bsnm{Cannity}, \binits{K.}},
\bauthor{\bsnm{Weiner}, \binits{R.}},
\bauthor{\bsnm{Hichenberg}, \binits{S.}},
\bauthor{\bsnm{Leon-Nastasi}, \binits{A.}},
\bauthor{\bsnm{Banerjee}, \binits{S.}},
\bauthor{\bsnm{Parker}, \binits{P.}}:
\batitle{General and unique communication skills challenges for advanced practice providers: A mixed-methods study}.
\bjtitle{Journal of the Advanced Practitioner in Oncology}
\bvolume{13}(\bissue{1}),
\bfpage{32}
(\byear{2022})
\end{barticle}
\endbibitem

\bibitem[\protect\citeauthoryear{Murugesu et~al.}{2022}]{murugesu2022challenges}
\begin{barticle}
\bauthor{\bsnm{Murugesu}, \binits{L.}},
\bauthor{\bsnm{Heijmans}, \binits{M.}},
\bauthor{\bsnm{Rademakers}, \binits{J.}},
\bauthor{\bsnm{Fransen}, \binits{M.P.}}:
\batitle{Challenges and solutions in communication with patients with low health literacy: Perspectives of healthcare providers}.
\bjtitle{PLoS One}
\bvolume{17}(\bissue{5}),
\bfpage{0267782}
(\byear{2022})
\end{barticle}
\endbibitem

\bibitem[\protect\citeauthoryear{Rehim et~al.}{2017}]{rehim2017tools}
\begin{barticle}
\bauthor{\bsnm{Rehim}, \binits{S.A.}},
\bauthor{\bsnm{DeMoor}, \binits{S.}},
\bauthor{\bsnm{Olmsted}, \binits{R.}},
\bauthor{\bsnm{Dent}, \binits{D.L.}},
\bauthor{\bsnm{Parker-Raley}, \binits{J.}}:
\batitle{Tools for assessment of communication skills of hospital action teams: a systematic review}.
\bjtitle{Journal of surgical education}
\bvolume{74}(\bissue{2}),
\bfpage{341}--\blpage{351}
(\byear{2017})
\end{barticle}
\endbibitem

\bibitem[\protect\citeauthoryear{Srinivasan et~al.}{2006}]{srinivasan2006connoisseurs}
\begin{barticle}
\bauthor{\bsnm{Srinivasan}, \binits{M.}},
\bauthor{\bsnm{Franks}, \binits{P.}},
\bauthor{\bsnm{Meredith}, \binits{L.S.}},
\bauthor{\bsnm{Fiscella}, \binits{K.}},
\bauthor{\bsnm{Epstein}, \binits{R.M.}},
\bauthor{\bsnm{Kravitz}, \binits{R.L.}}:
\batitle{Connoisseurs of care? unannounced standardized patients’ ratings of physicians}.
\bjtitle{Medical care}
\bvolume{44}(\bissue{12}),
\bfpage{1092}--\blpage{1098}
(\byear{2006})
\end{barticle}
\endbibitem

\end{thebibliography}
